\documentclass[journal]{IEEEtran}
\usepackage{cite}
\usepackage{amsfonts}
\usepackage{amssymb}
\usepackage{amsmath}
\usepackage{arydshln}  %
\usepackage{graphicx}
\usepackage{tikz}
\usepackage{setspace} %
\usepackage{algorithm}
\usepackage{algorithmic}

\usepackage{hyperref}

{

}

\RequirePackage{amsmath}

\relax

\newcommand{\R}{\mathbb{R}}

\renewcommand{\S}{\mathbb{S}}

\newtheorem{thm}{Theorem}
\newtheorem{defn}[thm]{Definition}
\newtheorem{lem}[thm]{Lemma}

\newtheorem{cor}[thm]{Corollary}
\newtheorem{rem}[thm]{Remark}
\newtheorem{prop}[thm]{Proposition}
\newtheorem{ex}[thm]{Example} %

\newcommand{\norm}[1]{\lVert{#1}\rVert}

\newcommand{\bmat}[1]{\begin{bmatrix}#1\end{bmatrix}}

\def\PDKprime{
\begin{tikzpicture}[scale=0.5]

\draw  (-2,3) rectangle (2,1); %
\draw  (-1.15,0) rectangle (1.4,-2); %
\draw  (-1,-5) rectangle (1.3,-7); %
\draw (4.125,-3) rectangle (2.936,-4); %
\draw (-2.875,-3) rectangle (-4.063,-4); %
\draw[dashed]  (4.25,3.5) rectangle (-4.24,-4.5); %

\node[draw, circle] (adder) at (3.5,-1) {+};

\draw[ >=stealth, thick] [->] (-2,2) -- (-3.5,2) -- (-3.5,-3); %
\draw[ >=stealth, thick] [->] (-3.5,-4) -- (-3.5,-6) -- (-1,-6); %
\draw[ >=stealth, thick] [->] (-3.5,-1) -- (-1.15,-1); %

\draw[ >=stealth, thick] [->] (1.3,-6) -- (3.5,-6) -- (3.5,-4); %
\draw[ >=stealth, thick] [->] (3.5,-3) -- (adder); %
\draw[ >=stealth, thick] [->] (adder) -- (3.5,2)--(2,2); %
\draw[ >=stealth, thick] [->] (1.4,-1) -- (adder); %
\node[ >=stealth, thick] [->] at (0,2) {$P$};
\node[ >=stealth, thick] [->] at (0.15,-1) {$\Di{m-1}$};
\node[ >=stealth, thick] [->] at (0.15,-6) {$\pK$};
\node[ >=stealth, thick] [->] at (-3.5,-3.5) {$\e{j_m}^T$};
\node[ >=stealth, thick] [->] at (3.5,-3.5) {$\e{i_m}$};

\node at (2.75,2.5) {$u$};
\node at (-2.75,2.5) {$y$};
\node at (-2,-5.5) {$y^\prime$};
\node at (2,-5.5) {$u^\prime$};
\node at (0,4) {$P_m$};
\end{tikzpicture}}

\def\LFTfeedback{
\begin{tikzpicture}[scale=.65]

\def\PBsx{-2} %
\def\PBsy{3} %
\def\PBex{2} %
\def\PBey{1} %

\def\Kkcx{0}   %
\def\Kkcy{-1}   %
\def\Kkxl{2} %
\def\Kkyl{2} %

\def\Pxk{1} %
\def\PxS{1} %

\def\kxP{1} %
\def\SxP{1.75} %

\def\DBy{0.5} %

\draw  (\PBsx,\PBsy) rectangle (\PBex,\PBey);   %
\node at (\PBsx*0.5+\PBex*0.5,\PBsy*0.5+\PBey*0.5) {$P$}; %

\draw  (\Kkcx-0.5*\Kkxl,\Kkcy+0.5*\Kkyl) rectangle 
(\Kkcx+0.5*\Kkxl,\Kkcy-0.5*\Kkyl); %

\node at (\Kkcx,\Kkcy) {$K$};

\draw[ >=stealth, thick] [->] (\PBsx,\PBsy*0.5+\PBey*0.5) -- (\PBsx-\Pxk,\PBsy*0.5+\PBey*0.5) -- (\PBsx-\Pxk,\Kkcy) -- (\Kkcx-0.5*\Kkxl,\Kkcy);
\node at (\PBsx-\Pxk*0.5,\PBsy*0.5+\PBey*0.5+.25) {$y$};
\node at (\PBex+\kxP+\SxP*0.75,\PBsy*0.5+\PBey*0.5+.25) {$r$};

\draw[ >=stealth, thick] [->] (\PBsx-\Pxk,\PBsy*0.5+\PBey*0.5) -- (\PBsx-\Pxk-\PxS,\PBsy*0.5+\PBey*0.5); %

\draw[dashed] (\PBsx-\Pxk-\PxS*0.5,\PBsy+\DBy) rectangle (\PBex+\kxP+\SxP*0.55,\Kkcy - \Kkyl*0.5 - \DBy ) ;
\node at (\PBsx*0.5+\PBex*0.5,\PBsy+\DBy+.25) {$\LFT{P}{K}$};

\node[draw, circle] (adder) at (\PBex+\kxP,\PBsy*0.5+\PBey*0.5) {+};

\draw[ >=stealth, thick] [->] (adder)--(\PBex,\PBsy*0.5+\PBey*0.5) ;

\draw[ >=stealth, thick] [->] (\Kkcx+0.5*\Kkxl,\Kkcy) -- (\PBex+\kxP,\Kkcy) -- (adder);

\draw[ >=stealth, thick] [->] (\PBex+\kxP+\SxP,\PBsy*0.5+\PBey*0.5) -- (adder);

\end{tikzpicture}}

\def\Gnested{
\begin{tikzpicture}[scale=.55]

\def\PBsx{-2} %
\def\PBsy{3} %
\def\PBex{2} %
\def\PBey{1} %

\def\Kkcx{0}   %
\def\Kkcy{-1}   %
\def\Kkxl{2} %
\def\Kkyl{2} %

\def\KScx{0}   %
\def\KScy{-4}   %
\def\KSxl{2} %
\def\KSyl{2} %

\def\Pxk{1} %
\def\PxS{1} %

\def\kxP{1} %
\def\SxP{1.75} %

\def\DBy{0.5} %

\draw  (\PBsx,\PBsy) rectangle (\PBex,\PBey);   %
\node at (\PBsx*0.5+\PBex*0.5,\PBsy*0.5+\PBey*0.5) {$\Gs{k}$}; %

\draw  (\Kkcx-0.5*\Kkxl,\Kkcy+0.5*\Kkyl) rectangle (\Kkcx+0.5*\Kkxl,\Kkcy-0.5*\Kkyl); %

\node at (\Kkcx,\Kkcy) {$\Kms{k}$};

\draw[ >=stealth, thick] [->] (\PBsx,\PBsy*0.5+\PBey*0.5) -- (\PBsx-\Pxk,\PBsy*0.5+\PBey*0.5) -- (\PBsx-\Pxk,\Kkcy) -- (\Kkcx-0.5*\Kkxl,\Kkcy);
\node at (\PBsx-\Pxk*0.5,\PBsy*0.5+\PBey*0.5+.25) {$y$};
\node at (\PBex+\kxP+\SxP-0.25,\PBsy*0.5+\PBey*0.5+.25) {$u$};

\draw[ >=stealth, thick] [->] (\PBsx-\Pxk,\PBsy*0.5+\PBey*0.5) -- (\PBsx-\Pxk-\PxS,\PBsy*0.5+\PBey*0.5);

\draw[dashed] (\PBsx-\Pxk-\PxS*0.5,\PBsy+\DBy) rectangle (\PBex+\kxP+\SxP*0.65,\Kkcy*0.5+\KScy*0.5) ;
\node at (\PBsx*0.5+\PBex*0.5,\PBsy+\DBy+.45) {$\Gs{k+1}$};

\node[draw, circle] (adder) at (\PBex+\kxP,\PBsy*0.5+\PBey*0.5) {+};

\draw[ >=stealth, thick] [->] (adder)--(\PBex,\PBsy*0.5+\PBey*0.5) ;

\draw[ >=stealth, thick] [->] (\Kkcx+0.5*\Kkxl,\Kkcy) -- (\PBex+\kxP,\Kkcy) -- (adder);

\draw[ >=stealth, thick] [->] (\PBex+\kxP+\SxP+0.25,\PBsy*0.5+\PBey*0.5) -- (adder);

\end{tikzpicture}}

\def\Pnested{
\begin{tikzpicture}[scale=.65]

\def\PBsx{-2} %
\def\PBsy{3} %
\def\PBex{2} %
\def\PBey{1} %

\def\Kkcx{0}   %
\def\Kkcy{-1}   %
\def\Kkxl{2} %
\def\Kkyl{2} %

\def\KScx{0}   %
\def\KScy{-4}   %
\def\KSxl{2} %
\def\KSyl{2} %

\def\Pxk{1} %
\def\PxS{1} %

\def\kxP{1} %
\def\SxP{1.25} %

\def\DBy{0.5} %

\draw  (\PBsx,\PBsy) rectangle (\PBex,\PBey);   %
\node at (\PBsx*0.5+\PBex*0.5,\PBsy*0.5+\PBey*0.5) {$P$}; %

\draw  (\Kkcx-0.5*\Kkxl,\Kkcy+0.5*\Kkyl) rectangle (\Kkcx+0.5*\Kkxl,\Kkcy-0.5*\Kkyl); %

\node at (\Kkcx,\Kkcy) {$\Ks{k}$};

\draw  (\KScx-0.5*\KSxl,\KScy+0.5*\KSyl) rectangle (\KScx+0.5*\KSxl,\KScy-0.5*\KSyl); %
\node at (\KScx,\KScy) {$\Ks{\star}$};

\draw[ >=stealth, thick] [->] (\PBsx,\PBsy*0.5+\PBey*0.5) -- (\PBsx-\Pxk,\PBsy*0.5+\PBey*0.5) -- (\PBsx-\Pxk,\Kkcy) -- (\Kkcx-0.5*\Kkxl,\Kkcy);
\node at (\PBsx-\Pxk*0.5,\PBsy*0.5+\PBey*0.5+.25) {$y$};
\node at (\PBex+.3750,\PBsy*0.5+\PBey*0.5+.25) {$u$};

\draw[ >=stealth, thick] [->] (\PBsx-\Pxk,\PBsy*0.5+\PBey*0.5) -- (\PBsx-\Pxk-\PxS,\PBsy*0.5+\PBey*0.5) -- (\PBsx-\Pxk-\PxS,\KScy) -- (\KScx-0.5*\KSxl,\KScy);

\draw[dashed] (\PBsx-\Pxk-\PxS*0.5,\PBsy+\DBy) rectangle (\PBex+\kxP+\SxP*0.65,\Kkcy*0.5+\KScy*0.5) ;
\node at (\PBsx*0.5+\PBex*0.5,\PBsy+\DBy+.35) {$\Pk{k}$};

\node[draw, circle] (adder) at (\PBex+\kxP,\PBsy*0.5+\PBey*0.5) {+};

\draw[ >=stealth, thick] [->] (adder)--(\PBex,\PBsy*0.5+\PBey*0.5) ;

\draw[ >=stealth, thick] [->] (\Kkcx+0.5*\Kkxl,\Kkcy) -- (\PBex+\kxP,\Kkcy) -- (adder);

\draw[ >=stealth, thick] [->] (\KScx+0.5*\KSxl,\KScy) -- (\PBex+\kxP+\SxP,\KScy)-- (\PBex+\kxP+\SxP,\PBsy*0.5+\PBey*0.5) -- (adder);

\end{tikzpicture}}

\def\PDplusV{

\begin{tikzpicture}[scale=0.55]

\draw  (-2,3) rectangle (2,1); %
\draw  (-1,0) rectangle (1,-2); %
\draw (4,-3) rectangle (3,-4); %
\draw (-3,-3) rectangle (-4,-4); %
\draw[dashed]  (4.25,3.5) rectangle (-4.24,-4.5); %

\node[draw, circle] (adder) at (3.5,-1) {+};

\draw[ >=stealth, thick] [->] (-2,2) -- (-3.5,2) -- (-3.5,-3); %
\draw[ >=stealth, thick] [->] (-3.5,-4) -- (-3.5,-6); %
\draw[ >=stealth, thick] [->] (-3.5,-1) -- (-1,-1); %

\draw[ >=stealth, thick] [->]  (3.5,-6) -- (3.5,-4); %
\draw[ >=stealth, thick] [->] (3.5,-3) -- (adder); %
\draw[ >=stealth, thick] [->] (adder) -- (3.5,2)--(2,2); %
\draw[ >=stealth, thick] [->] (1,-1) -- (adder); %
\node[ >=stealth, thick] [->] at (0,2) {$P$};
\node[ >=stealth, thick] [->] at (0,-1) {$\Dk$};
\node[ >=stealth, thick] [->] at (-3.5,-3.5) {\hspace{1.5pt}$\e{j}^T$};
\node[ >=stealth, thick] [->] at (3.5,-3.5) {\hspace{1.5pt}$\e{i}$};

\node at (2.75,2.5) {$u$};
\node at (-2.75,2.5) {$y$};
\node at (-4.00,-5.5) {$y^\prime$};
\node at (4,-5.5) {$u^\prime$};
\node at (0,4) {$P^+$};
\end{tikzpicture}}

\newcommand{\todo}[1]{%
}

\def\={~=~}

\def\Ap{A_P}
\def\Bp{B_P}
\def\Cp{C_P}
\def\Dp{D_P}
\newcommand{\Pk}[1]{P^{({#1})}}

\def\Ak{A_K}
\def\Bk{B_K}
\def\Ck{C_K}
\def\Dk{D_K}

\newcommand{\Ks}[1]{K^{({#1})}} %

\newcommand{\Aks}[1]{\Ak^{({#1})}}
\newcommand{\Bks}[1]{\Bk^{({#1})}}
\newcommand{\Cks}[1]{\Ck^{({#1})}}
\newcommand{\Dks}[1]{\Dk^{({#1})}}

\def\Acl{A_{\mathrm{CL}}}
\newcommand{\eig}[1]{\mathrm{eig}\left({#1}\right)}

\def\S{\mathcal{S}}
\def\Sc{\S_\mathrm{c}}
\def\Sd{\S_\mathrm{d}}

\newcommand{\Adm}[1]{\text{Adm}({#1})}

\def\sI{I}
\newcommand{\sIk}[1]{\sI^{({#1})}}

\def\T{\mathcal{T}}
\def\Ts{\T^{\mathrm{s}}}
\def\Td{\T^{\mathrm{d}}}
\newcommand{\Tspk}[1]{\T^{\text{s}+{#1}}}
\def\Tsp{\Tspk{1}}
\newcommand{\Tsk}[1]{\Tspk{{#1}}_{\sIk{{#1}}}}  %
\def\Tspo{\Tsp_{\isp,\jsp}}
\def\Kspo{K^{\mathrm{s}+1}}

\newcommand{\sK}[1]{\mathcal{K}_{{#1}}} %

\def\Pce{P^+} %
\def\isp{i}
\def\jsp{j}

\def\Apsp{A_{\Pce}}
\def\Bpsp{B_{\Pce}} %
\def\Cpsp{C_{\Pce}} %
\def\Dpsp{D_{\Pce}}

\newcommand\Eigp[1]{\text{eig}_+({#1})} %
\newcommand\Eigm[1]{\text{eig}_-({#1})} %
\def\Lamb{\zeta} %
\def\Lambt{\Lambda^\sim} %
\def\Lambp{\Lambda^\sim_{+}} %
\def\Lambm{\Lambda^\sim_{-}} %

\newcommand{\FM}[3]{ 
\Lambda\left({#1},{#2},{#3}\right)
}

\newcommand{\At}[1]{\tilde{A}_{{#1}} }
\newcommand{\Bt}[1]{\tilde{B}_{{#1}} }
\newcommand{\Ct}[1]{\tilde{C}_{{#1}} }

\def\tA{\tilde{A_K}}
\def\tB{\tilde{B_K}}
\def\tC{\tilde{C_K}}
\def\tD{\tilde{D_K}}

\def\Ah{\hat{A_K}}
\def\Bh{\hat{B_K}}
\def\Ch{\hat{C_K}}
\def\Dh{\hat{D_K}}

\newcommand{\Ats}[1]{\tA^{({#1})}}
\newcommand{\Bts}[1]{\tB^{({#1})}}
\newcommand{\Cts}[1]{\tC^{({#1})}}
\newcommand{\Dts}[1]{\tD^{({#1})}}

\newcommand{\Ahs}[1]{\Ah^{({#1})}}
\newcommand{\Bhs}[1]{\Bh^{({#1})}}
\newcommand{\Chs}[1]{\Ch^{({#1})}}
\newcommand{\Dhs}[1]{\Dh^{({#1})}}

\newcommand{\Khs}[1]{\hat{K}^{({#1})}} %
\newcommand{\Kts}[1]{\tilde{K}^{({#1})}} %
\newcommand{\eigCLl}[1]{\eig{\Acl(P,{#1})}}  %
\newcommand{\eigCLr}[1]{\eig{\Acl(\Pk{k},{#1})}} %
\newcommand{\Ptk}[1]{\tilde{P}^{({#1})}}
\newcommand{\Phk}[1]{\hat{P}^{({#1})}}
\newcommand{\eigCLrt}[1]{\eig{\Acl(\Ptk{k},{#1})}} %
\newcommand{\eigCLrh}[1]{\eig{\Acl(\Phk{k},{#1})}} %

\def\FMLHS{\FM{P}{\S}{\Tsk{k+1}}}
\def\FMRHS{\bigcap\limits_{\Ks{k}\in\S
\cap \Tsk{k}}  \FM{\Pk{k}}{\S}{\Tsp_{i, j}}}

\newcommand{\e}[1]{\mathbf{e}_{{#1}}}   %

\newcommand{\LFT}[2]{\Gamma({#1},{#2})}

\def\LHP{\mathbb{C}^{-}}
\def\RHP{\mathbb{C}^{+}}

\def\pA{A^\prime}
\def\pB{B^\prime}
\def\pC{C^\prime}

\def\pK{K^\prime}

\def\Pp{W}

\def\mK{K_m}
\def\mA{A_m^K}
\def\mB{B_m^K}
\def\mC{C_m^K}
\def\mD{D_m^K}

\def\matk{m^{(k)}}
\newcommand{\Kms}[1]{K_m^{({#1})}} %

\newcommand{\Gs}[1]{G^{({#1})}}
\newcommand{\nuk}[1]{\nu^{(#1)}}

\def\Ag{A_G}
\def\Bg{B_G}
\def\Cg{C_G}
\def\Dg{D_G}

\newcommand{\Di}[1]{D|_{({#1})}}
\def\Dshort{\breve{D}}

\newcommand{\Ags}[1]{\Ag^{({#1})}}
\newcommand{\Bgs}[1]{\Bg^{({#1})}}
\newcommand{\Cgs}[1]{\Cg^{({#1})}}
\newcommand{\Dgs}[1]{\Dg^{({#1})}}

\newcommand{\atk}[1]{{#1}^{(k)}} %

\def\Dkalg{D}  %
\def\Dkalgm{D|_{(m)}}

\def\Dkt{\tilde{D}_K}  %
\def\Pco{P_{\mathrm{co}}} %

\allowdisplaybreaks  %
\begin{document}
\title{Constructive Stabilization and Pole Placement \\ 
	by Arbitrary Decentralized Architectures}

\author{\phantom{}\hskip 25pt Alborz Alavian \qquad and\qquad Michael Rotkowitz
\thanks{A. Alavian is with the Department of Electrical and
  Computer Engineering,
  University of Maryland, College Park, MD 20742 USA,
  {\tt\small alavian@umd.edu}.} %
\thanks{M. C. Rotkowitz is with 
the Institute for Systems Research
 and 
 the Department of Electrical and Computer Engineering, 
  The University of Maryland, College Park, MD 20742 USA,
  {\tt\small mcrotk@umd.edu}.} %
\thanks{This material is based upon work supported by the National
  Science Foundation under Grant No. 1351674.}
}

\maketitle

\begin{abstract}

A seminal result in decentralized control is the development of fixed
modes by Wang and Davison in 1973 - that plant modes which cannot be moved
with a static decentralized controller cannot be moved by a dynamic one
either, and that the other modes which can be moved can be shifted to any
chosen location with arbitrary precision. 
These results were developed for perfectly decentralized,
or block diagonal, information structure, where each control input may
only depend on a single corresponding measurement. Furthermore, the
results were claimed after a preliminary step was demonstrated, omitting a
rigorous induction for each of these results, and the remaining task is
nontrivial. 

In this paper, we consider fixed modes for arbitrary information
structures, where certain control inputs may depend on some measurements
but not others. We provide a comprehensive proof that 
the modes which cannot be altered by a static controller with the 
given structure cannot be moved by a dynamic one either,
and that the modes which can
be altered by a static controller with the given structure can be moved by
a dynamic one to any chosen location with arbitrary precision, 
thus generalizing and solidifying
Wang and Davison's results.

This shows that a system can be stabilized by a linear time-invariant
controller with the given information structure as long as all of the
modes which are fixed with respect to that structure are in the left
half-plane; an algorithm for synthesizing such a stabilizing decentralized
controller is then distilled from the proof.

\end{abstract}

\IEEEpeerreviewmaketitle

\section{Introduction}
\label{sec:intro}
This paper is concerned with the stabilization of decentralized
control systems, for which certain controller inputs may depend on
some measurements but not others. This corresponds to finding a
stabilizing controller which satisfies a given sparsity constraint.
A special case of this, sometimes
referred to as \emph{perfectly decentralized control}, 
occurs when each control input may depend only on a single associated
measurement, which corresponds to finding a stabilizing controller
which is (block) diagonal. 

This special case is sometimes itself referred to as \emph{decentralized control},
particularly in the literature from a few decades ago.
This malleability or evolution of the definition has not only caused
some confusion, but has also resulted in some important results in the
field only being studied for this special case.

We assume that plants and controllers are finite-dimensional, linear
time-invariant (FDLTI), except for when we say otherwise.

A seminal result in decentralized control is the development of 
\emph{fixed modes} by Wang and Davison in 1973\cite{wang_1973}.
This paper studied FDLTI perfectly decentralized stabilization of
FDLTI systems.  
Its contributions can be broken into three main components - a
definition establishing the framework, and two subsequent results.
Fixed modes were defined as those modes of the plant which could not
be altered by any static perfectly
decentralized controller (that is, by any diagonal matrix).  
The first result was 
that these fixed modes could also not be 
altered by any dynamic perfectly decentralized controller; 
if you can't move it with a static diagonal controller, you can't move
it with a dynamic diagonal controller.
The second result
was that if a mode is not fixed, 
then it can be moved
arbitrarily close to any chosen location 
in the complex plane (provided that it has a complex conjugate pair if
it is not real).  
These can be taken together to state
that a system is stabilizable by a (dynamic) perfectly decentralized controller
if and only if all of its (static) fixed modes are in the left
half-plane (LHP).

When proving these results, it was shown that allowing one part of the
controller to be dynamic does not result in any fewer fixed modes than
a static controller, and then claimed that 
the first result
followed; that is, that a dynamic controller would not be able to move
any of the fixed modes. Similarly, it was shown that a single non-fixed mode could
be moved to any chosen location, and then claimed that 
the second result
followed; that is, that an arbitrary number of non-fixed
modes could be simultaneously moved to chosen locations by a single
controller.
Getting from these initial steps to a rigorous inductive argument,
however, is not trivial.

We seek to study these fundamental concepts for arbitrary information
structure, while developing robust notation and rigorous proofs, thus
placing the new and existing results on a sound mathematical footing.
 
We first introduce notation for fixed modes that allows it to vary
with information structure, as well as with the type of controllers
allowed (static, dynamic, linear, etc.). We then show that, for
arbitrary information structure, the fixed modes with respect to
dynamic controllers are the same as the fixed modes with respect to
static controllers.
Moreover, we provide a rigorous proof that the non-fixed modes can then be moved to
within an arbitrarily small distance of chosen (conjugate) locations, using a
dynamic LTI controller with the given structure,
thus extending and solidifying the seminal
results of Wang and Davison. 
The proof is constructive, and we lastly distill an explicit algorithm
for the stabilizing decentralized controller synthesis from the proof.

The obvious potential benefits of this are 
an increased understanding of decentralized stabilizability, and the
verification of important existing results.  It is also our hope that
the notation developed will be useful in further extending our
understanding of decentralized stabilizability to richer classes of
controllers for which the fixed modes may diminish relative to the
original static definition, particularly non-linear and/or
time-varying controllers
\cite{kobayashi1978controllability,anderson1981time,wang1982stabilization,gong1997stabilization}.
We further note that demonstrating the results of this paper directly
for arbitrary structure, as opposed to attempting to diagonalize the
problem and then prove the original perfectly decentralized results, would likely be
useful when other types of stability are required which are not
invariant under such transformations\todo{check/revisit},
 such as bounded-input bounded-output (BIBO) stability, 
though we currently focus on %
internal state stability. 
As an example of the diagonalization approach, readers are referred
to \cite{Pichai_Siljak_84}, where existence of a stabilizing 
controller under arbitrary information constraint 
has been demonstrated by transforming the problem 
into a diagonal one to which \cite{wang_1973} could be applied.
Furthermore,~\cite{Pichai_Siljak_84} demonstrates an analytical
test for determining structural fixed modes under arbitrary 
information constraint and shows its equivalence to a
graph-theoretical condition.

Dealing with the original structure is also
preferable since 
stabilizing controllers can be constructed without having to first
expand their size.
Finally, while the proofs in \cite{wang_1973}, (as well as
\cite{gong1997stabilization}), are constructive in nature, they do not
clearly lead to an explicit synthesis algorithm. 

Many of the ideas for the rigorous proof of the necessity of the fixed
mode condition were first presented
in~\cite{Alavian_Rotkowitz_mtns14}, while many of the ideas for
the rigorous proof of its sufficiency, along with the development of
the algorithm, were first presented in~\cite{alavian2014cdc},
before being refined and generalized here.

The organization of this paper is as follows. In Section~\ref{sec:prelims} we define
notation and preliminaries, including our definition of fixed modes
and the controller types that we will later need. In
Section~\ref{sec:review} 
we review and establish
some results for centralized controllers. 
In Section~\ref{sec:main}, we then state and
prove our main results in two parts; 
Section~\ref{sec:mainNecc} will prove the necessity of having 
fixed modes in the LHP 
for existence of a FDLTI stabilizing controller, 
and Section~\ref{sec:mainSuff} will prove the sufficiency of the aforementioned assumption.
In Section~\ref{sec:numex}, we give the
explicit computational algorithm, along with a numerical example,
followed by some concluding remarks in Section~\ref{sec:conc}.

\section{Preliminaries}
\label{sec:prelims}
We proceed with the following preliminary definitions.
Let $\Re(\cdot)$ denote the real part of any complex number. 
Define $\mathbb{C}$ to be the complex plane, 
$\LHP \triangleq \{\lambda \in \mathbb{C} ~|~ \Re(\lambda) < 0  \}$ 
to be the open left-half plane, and 
$\RHP \triangleq \mathbb{C}\setminus\LHP =
 \{\lambda \in \mathbb{C} ~|~ \Re(\lambda)\geq 0 \}$
to be the closed right-half plane.
Let $\e{i}$ denote the unit vector of all zeros 
except for the $i^\mathrm{th}$ element which is 1.
Note that the dimension of $\e{k}$ should be clear from the context 
and thus we suppress the explicit dimension of $\e{k}$ 
in the notation.
For a real matrix $A$, define the following norm:
\begin{equation*}
\norm{A}_\infty = \max_{i}\left(\sum\nolimits_{j} \lvert A_{ij} \rvert \right),
\end{equation*}
and let $B(\lambda_0,\epsilon) \triangleq 
\{ \lambda \in \mathbb{C} ~:~
\lvert \lambda - \lambda_0 \rvert < \epsilon \} $ 
denote the open $\epsilon$-ball around $\lambda_0$.

We consider an FDLTI plant 
$P(\sigma)$ 
(where $\sigma = s, z$ depending on whether we are considering
continuous or discrete-time cases; we use $\sigma$ for statements that
apply to both).
We assume that~$P$ has $n_u$ inputs, $n_y$ outputs, 
and a state-space representation of $P$ is given by $(\Ap,\Bp,\Cp,\Dp)$.
All controllers under consideration in this paper will also be
FDLTI. 

We impose information constraints on the controller
to encapsulate that each part of the controller may access certain sensor
measurements, but not others.
We define a set of admissible indices $\Adm{\S}$,
\todo{this is actually a bit backwards, as $\S$ should be a function of it}
such that $(i,j)\in\Adm{\S}$ if and only if controller $i$ is allowed
to access measurement $j$.
The information constraint is then denoted by the constraint $K\in\S$,
where $K_{ij}=0$ for all $(i,j)\notin\Adm{\S}$, for all $K\in\S$.

We will make use of the following specific sparsity patterns:
\begin{itemize}
\item $\Sc$: Centralized sparsity patterns, i.e., no sparsity
  constraints are imposed on the controller.
$\Adm{\S}=\{(i,j) ~~\forall~i,j\}$.

\item $\Sd$: Diagonal sparsity patterns, i.e., $K(\sigma)$ must be
  zero for all off-diagonal terms (for almost all $\sigma$).
$\Adm{\S}=\{(i,i) ~~\forall~i\}$.

\end{itemize}

We also define types of controllers that will help us to easily refer
to whether a controller $K$ is static, dynamic, or static 
for some elements but dynamic for others.
We will make use of the following controller types:

\begin{itemize}
\item $\Td$: Set of finite order dynamic controllers, i.e.,
  $\Ak,\Bk,\Ck,\Dk$ each are real matrices of compatible dimension.

\item $\Ts$: Set of static controllers, i.e., $\Ak, \Bk$, and $\Ck$ are
  all zero and only $\Dk$ could be non-zero.

\item $\Tsp_{\isp,\jsp}$: Set of controllers such that all of the
  elements of the controller are static except for the
  $(\isp,\jsp)^{\text{th}}$ element which could be dynamic; i.e., for
  all $(k,l) \neq (\isp,\jsp)$, we have ${K_{kl} \in \mathbb{R}}$, while
  $K_{\isp\jsp}$ may be a proper transfer function in $\sigma$. 
  This could be read as ``static plus one''.

\item $\Tspk{k}_\sI$: Set of controllers such that all the elements of
  controller are static except for $k$ indices in the set $\sI
  \triangleq 
\left\{ (i_1,j_1), \cdots , (i_k,j_k) \right\}$;
i.e., for all $(k,l) \notin \sI$, we have $K_{kl} \in\R$, while
 $K_{ij}$ is a proper transfer function
in~$\sigma$ for all $(i,j) \in \sI$. 
This could be read as ``static plus $k$''.
\end{itemize}

For any information structure~$\S$, 
let $a \triangleq \lvert \Adm{\S} \rvert$ be the number of 
admissible non-zero indices of the controller, and
let the tuple $\sI \triangleq \{(i_1,j_1),\cdots,(i_a,j_a) \}$ be any arbitrary
ordering of these admissible non-zero indices. %
For any
 $D \in \Ts\cap \S$,
 we define the %
sequence of matrices $\Di{m}\in\R^{n_u\times n_y}$, $m \in \{0,1,\cdots,a\}$ as:
\begin{equation} \label{eq:DmDef}
\Di{0} \triangleq 0,\qquad \Di{m} \triangleq \sum\limits_{l=1}^{m} 
\e{i_l}D_{i_l j_l} \e{j_l}^T
\quad \text{for}~m \in \{1,\cdots,a\}
\end{equation}
where $\e{i_l} \in \mathbb{R}^{n_u}$ and
$\e{j_l} \in \mathbb{R}^{n_y}$, for $l \in \{1,\cdots,a\}$.
This~$\Di{m}$ gives the static controller matrix with only the first~$m$ admissible indices.

The closed-loop has a state-space representation with dynamics matrix
denoted by $\Acl(P,K)$, given by:
\begin{equation} \label{eq:CL_general}        
\begin{aligned}
\Acl&(P, K) ~\triangleq\\
&\left(
\begin{array}{c  c}
\Ap + \Bp M\Dk\Cp & \Bp M\Ck \\ %
\Bk N\Cp & \Ak + \Bk\Dp M\Ck
\end{array}
\right),
\end{aligned}
\end{equation}
where $M\triangleq(I - \Dk\Dp)^{-1}$ and $N\triangleq(I - \Dp\Dk)^{-1}$.
We have $M\Dk = \Dk N$ and similarly $\Dp M = N\Dp$,
as well as
$N = I + \Dp M\Dk$.

\begin{figure} [htbp]
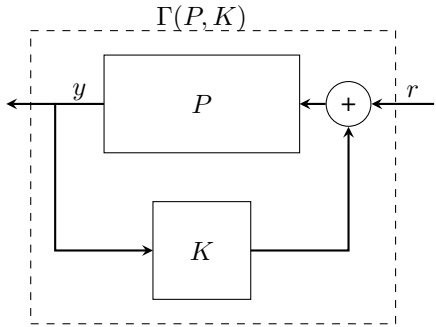
 
\begin{center}
\LFTfeedback
\end{center}
\caption{The map from reference inputs to outputs when $K$ is closed
around~$P$.} 
\label{fig:LFT}
\end{figure}

As illustrated in Figure \ref{fig:LFT},
let $\LFT{P}{K}$ denote 
the map from the reference inputs 
to the outputs of $P$
(i.e., from $r$ to $y$),
when $K$ is closed around $P$.
A state-space representation for $\LFT{P}{K}$ is given by:
\begin{equation} \label{eq:LFTss}
\LFT{P}{K} = \left[\begin{array}{c | c}
\Acl(P,K) & \begin{array}{c}
\Bp M \\ \Bk\Dp M
\end{array} \\ \hline
\begin{array}{c c}
N\Cp & \Dp M \Ck
\end{array} & \Dp M
\end{array}
\right]. %
\end{equation}

We have the following property of $\LFT{\cdot}{\cdot}$:
\begin{equation} \label{eq:LFTprop}
\LFT{\LFT{P}{K_1}}{K_2} ~=~ \LFT{P}{K_1+K_2},
\end{equation}
which can be verified by working out the state-space 
representation of both sides.

\begin{defn} \label{def:FM}
The set of fixed modes of a plant $P$ with respect to a sparsity pattern
$\S$ and a type $\T$, is defined to be:
\begin{equation*} \label{eq:FM_def}
\begin{aligned}
\FM{P}{\S}{\T} &\triangleq \left\{ \lambda \in \mathbb{C} \: | \: \lambda \in \eig{\Acl(P,K)} \:  \forall~K
 \in \S 
\cap \T   \right\}\\
&=\underset{K\in\S\cap\T}{\bigcap} \eig{\Acl(P,K)}.
\end{aligned}
\end{equation*}
\end{defn}
\begin{rem}
This reduces to the definition of fixed modes in~\cite{wang_1973} if
$\S=\Sd$ and $\T=\Ts$.
\end{rem}

For any FDLTI system $P$, 
denote 
its open-loop modes
by 
$\Lamb (P)=\eig{\Ap}$, 
and for each mode 
$\lambda \in \Lamb(P)$, 
let $\mu(\lambda,P)$ denote its multiplicity.
We can partition the open-loop modes as:
\begin{equation} \label{eq:eigPpart}
\Lamb (P) 
~=~ \FM{P}{\S}{\Ts} ~\cup~ \Lambt(P,\S,\Ts)
\end{equation}
where 
\begin{equation*} 
\Lambt(P,\S,\Ts) = \eig{\Ap} \setminus \FM{P}{\S}{\Ts} 
\end{equation*}
gives the non-fixed modes, which we then further partition as:
\begin{equation*}
\Lambt(P,\S,\Ts) ~=~ \Lambp(P,\S,\Ts) ~\cup~ \Lambm(P,\S,\Ts),
\end{equation*}
where
\begin{equation*} %
\begin{aligned}
\Lambp(P,\S,\Ts) &= \{\alpha\in\Lamb(P) \: | \: 
\Re(\alpha) \geq 0\}
 \setminus \FM{P}{\S}{\Ts} \\
&= \Lambt(P,\S,\Ts) ~\cap~ \RHP\\
&= \{\alpha_1,\cdots,\alpha_{\lvert \Lambp(P) \rvert} \}
\end{aligned}
\end{equation*}
are distinct unstable non-fixed open-loop eigenvalues of $P$, 
and 
\begin{equation*} %
\begin{aligned}
\Lambm(P,\S,\Ts) &= \{\beta\in\Lamb(P) \: | \: 
 \Re(\beta)<0 \} 
 \setminus \FM{P}{\S}{\Ts}\\ 
&= \Lambt(P,\S,\Ts) ~\cap~ \LHP\\
&= \{\beta_1,\cdots,\beta_{\lvert \Lambm(P) \rvert}   \}
\end{aligned}
\end{equation*}
are distinct stable non-fixed open-loop eigenvalues of $P$. 
We may suppress the dependence of these collections of eigenvalues on
some of their arguments when clear from context.
 
We note that one can adopt the notion of the multiset
to discriminate between copies of a mode with multiplicity greater than one.
This would have some conceptual advantages, but would unnecessarily
complicate some definitions and proofs, and so we maintain the use of
standard sets, while tracking the multiplicities of the modes which we
will want to move (the unstable non-fixed modes).
This is equally acceptable, provided that a fixed and a non-fixed mode
do not have the same value, which would require the non-fixed modes to
be defined as something other than the complement of those which are
fixed, as above (and multiset complementation could handle this aspect
nicely). Even that situation could not be problematic if we are
considering the complex plane as being split into an acceptable and an
unacceptable region, since such an overlap would either represent an
acceptable situation, or one which is fatal anyway.

Denote the total (with multiplicities) number of 
unstable non-fixed modes of a plant $P$ 
by %
\begin{equation*}
\nu(P) ~\triangleq \sum\limits_{\alpha \in  \Lambp(P) } \mu(\alpha,P). 
\end{equation*}

For a matrix $A$, 
we refer to the non-negative and negative eigenvalues respectively by
$\Eigp{A} \triangleq \eig{A} \cap \RHP$, 
and $\Eigm{A} \triangleq \eig{A}\cap\LHP $.
When~$\Eigm{\cdot}$, and~$\Eigp{\cdot}$ 
are applied on a general LTI system~$P$, 
with a slight abuse of notation, 
we mean the negative, and non-positive eigenvalues of the dynamic matrix of 
that system, 
i.e.,~$\Eigm{P}\triangleq\Eigm{\Ap}$, 
and~$\Eigp{P}\triangleq\Eigp{\Ap}$.

\section{Centralized Results}
\label{sec:review}
In this section we review and establish results on controllability,
observability, and fixed modes for centralized control of linear
time-invariant systems.
We begin with
Kalman canonical form with the help of the following lemma:

\begin{lem} \label{lem:Kalman_dc}
For every FDLTI plant $P$, 
there exists a similarity transformation matrix $T$ such that 
\def\arraystretch{1.2}
\begin{align}%
\label{eq:Kalman_dc}
\begin{bmatrix} 
T & 0 \\
0 & I
\end{bmatrix}
\vspace{3 mm}
\left[
\begin{array}{l | r}
\Ap & \Bp \\ \hline
\Cp & \Dp
\end{array}
\right]
\begin{bmatrix}
T^{-1} & 0 \\
0 & I
\end{bmatrix}
& \notag \\
 = 
\left[
\begin{array}{c c c c | c}
\At{11} & 0 & \At{13} & 0 & \Bt{1} \\
\At{21} & \At{22} & \At{23} & \At{24} & \Bt{2} \\
0 & 0 & \At{33} & 0 & 0 \\
0 & 0 & \At{43} & \At{44} & 0 \\ \hline
\Ct{1} & 0 & \Ct{2} & 0 & \Dp
\end{array}
\right].
\end{align}
In the above equation we have the following correspondence between
eigenvalues of $\At{ii}$ and modes of $P$:
\begin{itemize}
\item $\At{11}$: controllable and observable modes of $P$,

\item $\At{22}$: controllable and unobservable modes of $P$,

\item $\At{33}$: uncontrollable and observable modes of $P$,

\item $\At{44}$: uncontrollable and unobservable modes of $P$.
\end{itemize}

\end{lem}

\begin{IEEEproof}
See, for example, \cite{Kalman_62}.
\end{IEEEproof}

In order to reduce some of the notation,
we do not explicitly
show the dependence of $\At{ij},\Bt{i}, \Ct{j}$ on $\Ap,\Bp, \Cp,$ and
$T$, but it should be kept in mind that wherever we use 
Lemma~\ref{lem:Kalman_dc} on a system, 
the resulting $\tilde{(\cdot)}$
variables are function of that system's state-space matrices with its
respective Kalman similarity transformation matrix. 

The following lemma is useful in connecting centralized fixed modes 
with the familiar notion of controllability and observability.
It was shown for strictly proper plants in~\cite{Davison_Wang_73}; we
establish the following generalization before proceeding.
\begin{lem} \label{lem:DkSc}
Given a proper controllable and observable plant~$\Pco$, 
for almost any~$\Dk\in\Ts\cap\Sc$, we have that:
\begin{equation} \label{eq:lemDkSc}
\eig{\Acl(\Pco, \Dk)} \cap \eig{\Pco} ~=~ \varnothing.
\end{equation}
\end{lem}
\begin{IEEEproof}
For a strictly proper plant refer to~\cite[Theorem~2]{Davison_Wang_73}.
Given the proper plant~$\Pco$, %
consider the strictly proper part of it, namely ~$\Pco-D$.
Then, by~\cite[Theorem~2]{Davison_Wang_73} the set of static feedback gains~$\Dkt$
for which~$\eig{\Acl(\Pco-D, \Dkt)} \cap \eig{\Pco-D} \neq \varnothing$ constitute a 
finite union of hyperplanes in the ambient space, 
and hence almost any~$\Dkt\in\Ts\cap\Sc$ moves the open-loop eigenvalues of~$\Pco-D$.
If~$(I + \Dkt D)$ is invertible, 
then by the change of variable~$\Dk~=~(I + \Dkt D)^{-1}\Dkt$,
we have:
\begin{equation*}
\Acl(\Pco-D, \Dkt)~=~ \Acl(\Pco, \Dk).
\end{equation*}
The proof would be finished if~$(I + \Dkt D)$ is invertible for almost any~$\Dkt$.
This can be seen as~$\det(I + \Dkt D) = 0$ is a non-trivial polynomial in~$\Dkt$ 
(choosing~$\Dkt=0$ would yield non-zero determinant), 
and hence the set of~$\Dkt$ for which~$\det(I + \Dkt D) = 0$ 
is a set with dimension less than the ambient space and has zero measure.
\end{IEEEproof}
Next we state the following result regarding fixed modes with respect
to a centralized sparsity pattern $\Sc$, 
which tells us that the fixed modes of a plant 
with respect to a centralized information structure
are precisely its uncontrollable or unobservable modes.
\todo{review what Thm2 of D\&Y actually says}
\begin{lem}%
\label{lem:FM_ScTs}
For any FDLTI plant $P$,
\begin{equation*} \label{eq:lemFMScTs}
\FM{P}{\Sc}{\Ts} = \bigcup\limits_{i=2,3,4} \eig{ \At{ii}},
\end{equation*}
where $\At{ii}$ are the blocks in the Kalman canonical decomposition
of plant $P$, such that the fixed modes are the union of uncontrollable or
unobservable modes of $P$.
\end{lem}
\begin{IEEEproof}
Denote the controllable and observable part of~$P$ by 
$\Pco \triangleq \Ct{1}(sI - \At{11})^{-1}\Bt{1} + \Dp$.
We first establish that for any arbitrary~$\Dk \in \Sc\cap\Ts$ that is closed around~$P$, we have:
\begin{equation} \label{eq:lemDkeig}
\eig{\Acl(P, \Dk)}=\eig{\Acl(\Pco, \Dk)} \cup  (\bigcup\limits_{i=2,3,4} \eig{ \At{ii}}).
\end{equation}
To see this, apply the similarity transformation~$T$ given in 
Lemma~\ref{lem:Kalman_dc} on~$\Acl(P, \Dk)$.
Then~$T\Acl(P, \Dk)T^{-1}$ would only differ 
in blocks~$\At{11}, \At{21}, \At{13}$, and~$\At{23}$ 
compared to the open-loop~$\At{}$ in~\eqref{eq:Kalman_dc}.
This leaves the structure of~$\At{}$ unchanged, and renders~\eqref{eq:lemDkeig}.

For any~$\Dk\in\Sc\cap\Ts$, and for~$i=2, 3, 4$, we then have:
\begin{equation*}
\eig{\At{ii}}\subseteq\eig{T\Acl(P, \Dk)T^{-1}}=\eig{\Acl(P, \Dk)}, 
\end{equation*}
and so 
$\bigcup\limits_{i=2,3,4} \eig{ \At{ii}} \subseteq \FM{P}{\Sc}{\Ts}.$

For any remaining modes of~$P$, i.e., $\lambda\in\eig{\At{11}}$,
it follows from~\eqref{eq:lemDkeig} and Lemma~\ref{lem:DkSc}
that there exists a static controller~$\Dk\in\Sc\cap\Ts$ 
such that $\lambda\notin \eig{\Acl(P, \Dk)}$,
and so $\lambda\notin \FM{P}{\Sc}{\Ts}.$
\end{IEEEproof}

\begin{rem} \label{rem:almostanyDk}
Due to Lemma~\ref{lem:DkSc} and Lemma~\ref{lem:FM_ScTs},
almost any randomly chosen~$\Dk\in\Ts\cap\Sc$ moves all the open-loop modes of~$P$, 
except those of~$\FM{P}{\Sc}{\Ts}$. 
\end{rem}

Also\todo{edit, inc review what Lem3 of W\&D (and Thm2 of B\&P) actually says and proves} 
to make this paper sufficiently self-contained we use our
notation to restate the following result, which tells us that the
fixed modes of a plant with centralized information structure are the
same with respect to static or dynamic control.
\begin{thm} %
\label{thm:centralized}
Given an FDLTI plant $P$,
\todo{for this and previous lemma, qualify 'plant P', perhaps FDLTI
  plant P}
\begin{equation*} \label{eq:FMScTs_eq_Td}
\FM{P}{\Sc}{\Ts} = \FM{P}{\Sc}{\Td}.
\end{equation*}
\end{thm}
\begin{IEEEproof}
The $\supseteq$ inclusion
follows immediately since 
\mbox{$\Ts\subseteq\Td$.}

We now need to show that
$\FM{P}{\Sc}{\Ts} \subseteq \FM{P}{\Sc}{\Td}$; 
using Lemma~\ref{lem:FM_ScTs},
we can achieve this by showing that
$\bigcup\limits_{i=2,3,4} \eig{ \At{ii}} \subseteq \FM{P}{\Sc}{\Td},$
which can be achieved by showing that
$\bigcup\limits_{i=2,3,4} \eig{ \At{ii}} \subseteq \eig{\Acl(P,K)}$
for arbitrary $K\in\Sc\cap\Td$.

Given an arbitrary $K\in\Sc\cap\Td$, and letting $T$ be the similarity
transformation matrix from Lemma~\ref{lem:Kalman_dc},
we can then apply \eqref{eq:Kalman_dc} to \eqref{eq:CL_general}
to get
\begin{equation*} 
\begin{aligned}
&\begin{pmatrix}
T & 0 \\
0 & I
\end{pmatrix}
\Acl(P,K)
\begin{pmatrix}
T^{-1} & 0 \\
0 & I
\end{pmatrix}
\\ \vspace{3 mm}
&\hspace{15pt}=~\left(
\begin{array}{c  c}
\At{} + \Bt{}M\Dk\Ct{} & \Bt{} M\Ck \\ %
\Bk N\Ct{} & \Ak + \Bk\Dp M\Ck
\end{array}
\right)
\\ \vspace{3 mm}
&\hspace{15pt}=~\def\arraystretch{1.3}
\left(
\begin{array}{c c c c : c}
* & 0 & * & 0 & \Bt{1}M\Ck \\
* & \At{22} & * & * & \Bt{2}M\Ck \\
0 & 0 & \At{33} & 0 & 0 \\
0 & 0 & * & \At{44} & 0 \\ \hdashline
\Bk N\Ct{1} & 0 & \Bk N\Ct{2} & 0 & *
\end{array}
\right)
\end{aligned}
\end{equation*}
where we have let $(\At{},\Bt{},\Ct{},\Dp)$ give the blocks of
\eqref{eq:Kalman_dc}.

If we apply another similarity transformation which swaps the
first/second and third/fifth row and column blocks, the result is an
upper block triangular matrix for which the eigenvalues clearly
include those of $\At{22}$, $\At{33}$, and $\At{44}$, 
as desired.
\end{IEEEproof}

\section{Main Result}
\label{sec:main}
We will generalize the result of \cite{wang_1973} 
to arbitrary information structures,
and provide a comprehensive proof in this section. 
Section~\ref{sec:mainNecc} will establish the invariance 
of fixed modes with respect to static and dynamic controllers,
thereby demonstrating the necessity of having all of the fixed modes
in the LHP for decentralized stabilizability,
and Section~\ref{sec:mainSuff} will give a constructive proof for 
obtaining a stabilizing controller when all of the fixed modes of 
$P$ are in the LHP,
thereby demonstrating the sufficiency.

\subsection{Invariance of fixed modes}
\label{sec:mainNecc}
We will show in this subsection that for any arbitrary sparsity pattern $\S$, 
 the set of fixed modes with respect to static controllers is the
same as the set of fixed modes with respect to dynamic controllers.
\todo{provide brief outline?}

\begin{figure} [htbp]
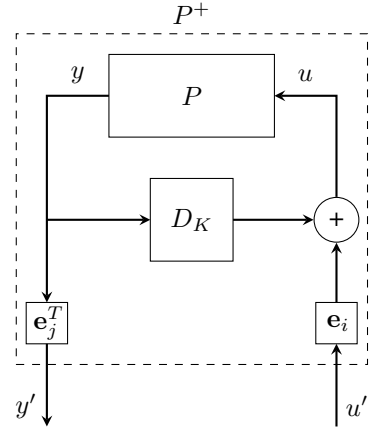

\begin{center}
\PDplusV
\end{center}
\caption{$\Pce$ is the SISO map from $u^\prime$ to $y^\prime$.
}
\label{fig:PDpl}
\end{figure}

We first state a lemma which is unsurprising but which will be helpful.
This lemma states that if $\lambda$ is a fixed
mode of a system with respect to static controllers and sparsity pattern~$\S$,
then after closing the loop with an arbitrary matrix $\Dk \in \S$, if we
further allow only one of the static elements of the controller
to vary, then $\lambda$ will
remain as a fixed mode.
Given any matrix $\Dk\in\S$, and any $(\isp,\jsp) \in\Adm{\S}$,
define $\Pce(\Dk)$, %
as illustrated in Figure \ref{fig:PDpl}, as:
\todo{note or capture the dependence of $\Pce$ on $\Dk$ and/or the index?}
\begin{equation*} \label{eq:PDpldef}
\Pce(\Dk) %
~\overset{\triangle}{=}~ 
\e{j}^T \LFT{P}{\Dk} \e{i} ~=~ 
\left[
\begin{array}{c | c}
\Apsp %
& \Bpsp \\ \hline
\Cpsp & \Dpsp
\end{array}
\right],
\end{equation*}
where 
$\Apsp\triangleq\Acl(P,\Dk)=\Ap+\Bp M\Dk\Cp$,
$\Bpsp\triangleq\Bp M\e{i}$, 
$\Cpsp\triangleq\e{j}^TN\Cp$, 
and ~$\Dpsp \triangleq \e{j}^T\Dp M\e{i}$.
\todo{thought it better to unclutter this for now}
We note that this notation suppresses the dependence of $\Pce$ on the
particular choice of %
 the admissible index pair.
\begin{onehalfspace}
\begin{lem} \label{lem:FMP_eq_FM_ABDkC}
Given any matrix $\Dk\in\S\cap\Ts$, 
and any $(\isp,\jsp) \in\Adm{\S}$,
 if $\lambda \in \FM{P}{\S}{\Ts}$, then $\lambda \in
\FM{\Pce(\Dk)}{\Sc}{\Ts}$, i.e., 
$\FM{P}{\S}{\Ts} \subseteq \FM{\Pce(\Dk)}{\Sc}{\Ts}$.
\end{lem}
\end{onehalfspace}
\begin{IEEEproof}
Suppose that $\lambda\in\FM{P}{\S}{\Ts}$.
\todo{I cleaned this proof up; there was no need for contradiction.}

For an arbitrary real scalar $V$, we have:
\begin{equation} \label{eq:pf_LEM_moveD}
\begin{aligned}
\Acl(\Pce(\Dk), V) ~&=~ \Acl(\e{j}^T \LFT{P}{\Dk} \e{i}, V) 
\\ ~&=~  \Acl(\LFT{P}{\Dk}, \e{i}V\e{j}^T)
\\ ~&\overset{\eqref{eq:LFTprop}}{=}~\Acl(P, \Dk + \e{i}V\e{j}^T)
\\ ~&=~\Acl(P, \Dk^V),
\end{aligned}
\end{equation}
where we have defined
$\Dk^V \triangleq \Dk + \e{i}V\e{j}^T$
as the static controller which is now effectively being closed around
the plant. 
Since we clearly have $\Dk^V\in\S\cap\Ts$ and 
since $\lambda\in\FM{P}{\S}{\Ts}$,
it follows that $\lambda\in\eig{\Acl(\Pce(\Dk),V)}$.
Since $V$ was arbitrary, we have
$\lambda\in\FM{\Pce(\Dk)}{\Sc}{\Ts}$.
\end{IEEEproof}

\todo{move fig?}
Next, we relate fixed modes with respect to static controllers 
to those where only one of the admissible elements %
is allowed to be dynamic; that is, to ``static plus one''
controllers. 
The lemma will prove useful because closing such a controller around
the plant is equivalent to interconnecting a SISO dynamic controller
with $\Pce$, and we can then leverage our knowledge of centralized
controllers.
\todo{wondering now if the whole thing can be made much simpler still. 
If we know that static/dynamic give same centralized FMs, then closing
static around $\Pce$ immediately gives same as dynamic gives what we
need for s+1}
This result will be the
foundation of the induction that we want to use later on.
The outline\todo{still true? perhaps just 'some steps'?} 
of the proof is similar %
to that of~\cite[Proposition~1]{wang_1973}.

\begin{thm} \label{thm:FMs1eq}
For any sparsity pattern $\S$, and any arbitrarily fixed indices $(\isp,\jsp) \in \Adm{\S}$:
\begin{equation*} \label{eq:FMs1eq}
\FM{P}{\S}{\Ts} = \FM{P}{\S}{\Tspo}.
\end{equation*}
\end{thm}
\vskip 3pt
\begin{IEEEproof}
The $\supseteq$ inclusion 
follows immediately since~\mbox{$\Ts\subseteq\Tspo$.}

\def\Ksiso{k^\mathrm{d}}
We now need to show that
$\FM{P}{\S}{\Ts} \subseteq \FM{P}{\S}{\Tspo}$.
We have:
\begin{align*}
\FM{P}{\S}{\Ts} 
&\overset{\text{Lem.\ref{lem:FMP_eq_FM_ABDkC}}}{\subseteq}
\bigcap\limits_{\Dk \in \S\cap\Ts}
\FM{\Pce(\Dk)}{\Sc}{\Ts} %
\\ \vspace{5 mm}
&\overset{\text{Thm.\ref{thm:centralized}}}{=} 
\bigcap\limits_{\Dk \in \S\cap\Ts}
\FM{\Pce(\Dk)}{\Sc}{\Td} %
\\ \vspace{5 mm}
&~=~\bigcap\limits_{\Dk \in \S\cap\Ts}
~\bigcap_{\Ksiso\in\Td}
\eig{\Acl(\Pce(\Dk),\Ksiso)}
\\ \vspace{5 mm}
&~=~\bigcap\limits_{\Dk \in \S\cap\Ts}
~\bigcap_{\Ksiso\in\Td}
\eig{\LFT{\LFT{P}{\Dk}}{\e{\isp}\Ksiso\e{\jsp}^T}}
\\ \vspace{5 mm}
&~\overset{\eqref{eq:LFTprop}}{=}
\bigcap\limits_{\Dk \in \S\cap\Ts}
~\bigcap_{\Ksiso\in\Td}
\eig{\LFT{P}{\Dk+\e{\isp}\Ksiso\e{\jsp}^T}}
\\ \vspace{5 mm}
&~=~
~\bigcap_{\Kspo\in\S\cap\Tspo}
\eig{\LFT{P}{\Kspo}}
\\ \vspace{5 mm}
&~=~\FM{P}{\S}{\Tspo}
\end{align*}
where the penultimate equality follows since
$(\S\cap\Ts)+\e{\isp}\Td\e{\jsp}^T=\S\cap\Tspo$,
and this completes the proof.
\end{IEEEproof}

\todo{remark or discussion on this equality and the failure of its generalization}
We note that it was this result, showing that modes which are fixed
with respect to static controllers are still fixed with respect to ``static plus one''
controllers,
 that was established for $\S=\Sd$ in
\cite{wang_1973}, and at which point Theorem~\ref{thm:statdynFM} was claimed
to hold true.
We will now show how to extend this result to show that modes which
are fixed with respect to controllers with any given number of dynamic
indices; that is, with respect to ``static plus $k$'' controllers, are
still fixed when an additional index is allowed to become dynamic;
that is, with respect to ``static plus $k+1$'' controllers.
The main result will indeed follow once that has been established.

We will proceed with the following definitions. 
Let $\Ks{k}(\sigma)$ be the controller after $k$ steps, with $k$ of
its indices allowed to be dynamic, and define
$\sIk{k}\triangleq
\left\{(i_1,j_1),\cdots,(i_k,j_k) \right\} \subset \Adm{\S}$
as the set of such indices where $\Ks{k}(\sigma)$ is allowed to be
dynamic,
such that
$\Ks{k}\in \Tsk{k} \cap \S$.
 Also let $(\Aks{k},\Bks{k},\Cks{k},\Dks{k})$ be a state-space representation for $\Ks{k}(\sigma)$. 
 
\begin{figure} [htbp]
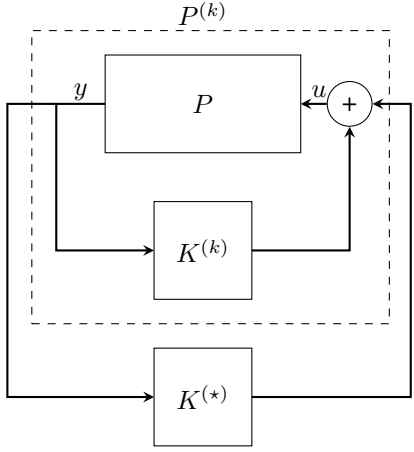

\begin{center}
\Pnested
\end{center}
\caption{Plant $\Pk{k}$ and its respective controller $\Ks{\star}$.} 
\label{fig:Pnested}
\end{figure}

 Define $\Pk{k}(\sigma)$, illustrated in Figure \ref{fig:Pnested}, 
by closing $\Ks{k}(\sigma)$ around $P(\sigma)$ in such a way 
 that the outputs of $\Pk{k}$ are the same as the outputs of $P$, 
 and such that the inputs of $\Pk{k}$ are added to the outputs of
 $\Ks{k}$ and fed into~$P$.

A state-space representation for  $\Pk{k}(\sigma)$ is given by
\begin{equation}\label{eq:Pk}
\Pk{k} \triangleq \LFT{P}{\Ks{k}},
\end{equation}
i.e., by replacing $(\Ak, \Bk, \Ck, \Dk, M)$ with
$(\atk{\Ak}, \atk{\Bk}, \atk{\Ck}, \atk{\Dk}, \atk{M})$ 
in \eqref{eq:LFTss}, 
where $\atk{M} = (I - \atk{\Dk}\Dp)^{-1}$.

We prove one remaining lemma before our main inductive step.
This lemma relates the modes which are fixed when closing controllers
with $k+1$ dynamic elements around the plant, to the modes which are
fixed when first closing controllers with $k$ dynamic elements around
the plant, and then closing a controller with an additional dynamic
element around the resulting plant, as in Figure~\ref{fig:Pnested}.
This will allow us to use our result relating static and  ``static plus one''
controllers to make conclusions relating ``static plus $k$'' and
``static plus $k+1$'' controllers.

\begin{rem}\label{rem:decRealizability}
We used the fact that $(\S\cap\Ts) +  \e{i}\Td\e{j}^T  = \S\cap\Tspo$;
that is, that adding static controllers and a dynamic element is
equivalent to taking all of the "static plus one" controllers,
at the end of the proof of Theorem~\ref{thm:FMs1eq}.
If this could be extended to state that 
$(\S\cap\Tsk{k}) + \Tspo = \S\cap\Tspk{k+1}_{\sIk{k} \cup (i, j)}$;
that is, that adding "static plus $k$" controllers and "static plus
one" controllers is equivalent to taking all of the "static plus
$k+1$" controllers, then Theorem~\ref{thm:FMsksub} would follow
similarly and easily, and the upcoming lemma would be trivial and
unnecessary.
It is not clear, however, that a "static plus $k+1$" controller can
always be decomposed in that manner.
We thus first introduce the following lemma, which states that,
regardless of whether those two sets are the same, the modes which
remain fixed as the controller varies over them are indeed identical.
\end{rem}

\begin{lem} \label{lem:indCLeq}
\begin{onehalfspace}
Given a set of indices $\sIk{k}\subset \Adm{\S}$, 
an additional index pair $(i, j) \in\Adm{\S} \setminus \sIk{k}$, 
set~$\sIk{k+1}$ to be~$\sIk{k+1} \triangleq \sIk{k} \cup (i, j)$,
and let $\Pk{k}$ be as in~\eqref{eq:Pk}, then we have:
\begin{equation} \label{eq:indCLeq}
\FM{P}{\S}{\Tsk{k+1}} =
\bigcap\limits_{\Ks{k}\in\S
\cap \Tsk{k}}
 \FM{\Pk{k}}{\S}{\Tspk{1}_{i, j}}.
\end{equation}
\end{onehalfspace}
\end{lem} 

\begin{IEEEproof}
For ease of notation, when the controllers are unambiguous such that
we can suppress the dependency upon them, define $\Acl^\text{LHS} = \Acl(P,\Ks{k+1})$ and
$\Acl^\text{RHS} = \Acl(\Pk{k},\Ks{\star})$ to be the closed-loop dynamics
matrices arising on each side of the equation for given controllers.
Also let 
$\sK{\text{LHS}} \triangleq \{ \Ks{k+1} | \Ks{k+1}  \in  \Tsk{k+1} \cap \S \}$, and 
$\sK{\text{RHS}} \triangleq \{(\Ks{k},\Ks{\star}) | \Ks{k} \in
\Tsk{k} \cap \S, \: \Ks{\star} \in \Tsp_{i, j} \cap \S\}$ 
give the sets of controllers that must be considered on
each side, such that the LHS can be abbreviated as
$\bigcap\limits_{ \sK{\text{LHS}}} \eig{\Acl^\text{LHS}}$, and the RHS
can be abbreviated as $\bigcap\limits_{\sK{\text{RHS}} } \eig{\Acl^\text{RHS}}$.
\vskip 1pt

 First we prove the $\subseteq$ part by showing that for every admissible $\Ks{\star}(\sigma)$, 
i.e., $\Ks{\star}\in\S   
\cap \Tsp_{i, j}$,  
and admissible  $\Ks{k}(\sigma)$
in RHS, there exist a $\Ks{k+1}(\sigma)$ in LHS such that
$\Acl^\text{RHS} = \Acl^\text{LHS}$.
To see this observe that:
\begin{equation*}
\LFT{\LFT{P}{\Ks{k}}}{\Ks{\star}} 
\overset{\eqref{eq:LFTprop}}{=}
\LFT{P}{\Ks{k} + \Ks{\star}}.
\end{equation*}
\begin{onehalfspace}Thus we choose 
$\Ks{k+1}(\sigma)  = \Ks{k}(\sigma) + \Ks{\star}(\sigma)$.
This $\Ks{k+1}$ is  admissible because it has only one further 
dynamic element at position $(i, j) \in 
\Adm{\S}$, and thus is in $\Tsk{k+1}$. Hence for every 
admissible $(\Ks{k},\Ks{\star})$, there exists an admissible 
$\Ks{k+1} \in \sK{\text{LHS}} $ constructed as above such that 
$\Acl^\text{LHS} = \Acl^\text{RHS}$, and so
$\bigcap\limits_{ \sK{\text{LHS}}} \eig{\Acl^\text{LHS}} \subseteq \bigcap\limits_{\sK{\text{RHS}} } \eig{\Acl^\text{RHS}}$.
\end{onehalfspace}

We will prove the $\supseteq$ part by contradiction. 
Assume that \eqref{eq:indCLeq} does not hold, 
and thus that there
exists a~$\lambda$ such that \mbox{$\lambda \in \FMRHS$,} 
but \mbox{$\lambda \notin \FMLHS$}.  Then %
we have:
\begin{subequations} \label{eq:forexits}
\begin{align} 
&\lambda\in\eigCLr{\Ks{\star}} ~\forall~  (\Ks{k},\Ks{\star}) \in \sK{\text{RHS}},  \label{eq:pf_THMfor} \\ %
 &\exists \: \Ks{k+1} \in \sK{\text{LHS}} \;   \text{ s.t.} \; \;  \; \lambda \notin \eigCLl{\Ks{k+1}}. \label{eq:pf_THMexists}
\end{align}
\end{subequations}

Starting with $\Ks{k+1}$ from \eqref{eq:pf_THMexists}, 
we will show that we can then construct 
a $\Ks{k}$ and $\Ks{\star}$ to 
 falsify \eqref{eq:pf_THMfor}.
\todo{subequations? onehalfspace? weird spacing?}

Based on $\Ks{k+1}$ in \eqref{eq:pf_THMexists}, 
we let $\Kts{\star}$ be the dynamic part of the final dynamic index by
defining $\Kts{\star} = (\Ats{\star},\Bts{\star} ,\Cts{\star} 
,\Dts{\star})$ as: %
\begin{equation*} \label{eq:Ktstar}
\begin{array}{l l} \vspace{2 mm}
\Ats{\star} &= \Aks{k+1}, \\ \vspace{2 mm}
\Bts{\star} &= \Bks{k+1}\e{j}\e{j}^T \\ &=
\begin{bmatrix}
0 & \cdots & B_{K,j}^{(k+1)} & \cdots & 0
\end{bmatrix}, \\ \vspace{2 mm}
\Cts{\star} &= \e{i}\e{i}^T\Cks{k+1} 
\\ &= \begin{bmatrix}
0 & \cdots & (C_{K,i}^{(k+1)})^T & \cdots & 0 
\end{bmatrix}^T,  \\ \vspace{2 mm}
\Dts{\star} &= 0,
\end{array}
\end{equation*}
i.e., $\Bts{\star}$ is of the same dimension as 
$\Bks{k+1}$ with all its columns being zero except the $j$-th 
column, and $\Cts{\star}$ is of the same dimension as $\Cks{k+1}$ with 
all of its rows being zero except the $i$-th row. Then define 
$\Kts{k} \triangleq \Ks{k+1} - \Kts{\star}$, thus a state-space 
representation for $\Kts{k}$ is:
\begin{equation*} \label{eq:Ktk}
\begin{array}{l l} \vspace{2 mm}
\Ats{k} = \text{diag}( \Aks{k+1},\Aks{k+1}), \\ \vspace{2 mm}
\Bts{k} = \begin{bmatrix}
(\Bks{k+1})^T & (\Bts{\star})^T
\end{bmatrix}^T, \\ \vspace{2 mm}
\Cts{k} = \begin{bmatrix}
\Cks{k+1} & -\Cts{\star}
\end{bmatrix},  \\ %
\Dts{k} = \Dks{k+1}.
\end{array}
\end{equation*}
Construct $\Ptk{k}$ in the same way as illustrated in Figure \ref{fig:Pnested} %
by closing $\Kts{k}$ around $P$. Now if we use the following similarity transformation $T$ on $\Acl(\Ptk{k},\Kts{\star})$,
\begin{equation*}
T = \begin{bmatrix}
0 & 0 & I & 0 \\
I & 0 & 0 & 0 \\
0 & I & 0 & 0 \\
0 & 0 & -I & I 
\end{bmatrix},
\end{equation*}
then $T \Acl(\Ptk{k},\Kts{\star}) T^{-1}$
 results in an upper block triangular matrix with blocks
  $\Aks{k+1}$, $\Acl(P,\Ks{k+1})$, and  $\Aks{k+1}$, 
indicating that:
\begin{equation} \label{eq:eigsub}
\begin{aligned}
&\eigCLrt{\Kts{\star}} = \\
&\hskip 40pt \eigCLl{\Ks{k+1}} \cup \eig{\Aks{k+1}}.
\end{aligned}
\end{equation}
Thus \eqref{eq:forexits} 
can only be satisfied if:
\begin{equation} \label{eq:lambdainA}
\lambda \in \eig{\Aks{k+1}}.
\end{equation}
We have shown that the only way to have an eigenvalue
 which is not on the LHS (when $\Ks{k+1}$ is closed around the plant)
 but which is on the RHS (when $\Kts{\star}$ and $\Kts{k}$ are then
 constructed as above), is if it comes from the dynamics matrix of
 $\Ks{k+1}$. 
We will now finish the proof by showing that if this is the case, we
can make a small perturbation to the matrix such that it no longer has
this eigenvalue, thus removing it from the RHS, while it is still not
a closed-loop eigenvalue on the LHS.

Construct $\Khs{k+1}$ by perturbing the $A$ matrix of $\Ks{k+1}$;
that is, $\Khs{k+1}$ is defined by:
\begin{equation*} \label{eq:Khstardef}
\begin{array}{l l l l}
\Ahs{k+1} = & \Ats{k+1} + \epsilon I, & \Bhs{k+1} =& \Bts{k+1}, \\
\Chs{k+1} =& \Cts{k+1}, & \Dhs{k+1} =& \Dts{k+1}. \\
\end{array}
\end{equation*} 
For sufficiently small $\epsilon$ this yields
\begin{equation} \label{eq:lambdanotinA} 
\lambda \notin \eig{\Ahs{k+1}}.
\end{equation}
Using the same steps as before
to construct $\Khs{\star}$ and $\Khs{k}$ 
results in $\epsilon I$ also being added to $\Ats{\star}$ and $\Ats{k}$.
Then using the same similarity transformation $T$ used 
to derive \eqref{eq:eigsub}, we have
 \begin{equation} \label{eq:eigsubh}
\begin{aligned}
& \eigCLrh{\Khs{\star}} = \\
&\hskip 40pt  \eigCLl{\Khs{k+1}} \cup \eig{\Ahs{k+1}},
\end{aligned}
 \end{equation}
where $\Phk{k}$ is constructed
by closing $\Khs{k}$ around $P$,
as illustrated for the unperturbed systems in Figure \ref{fig:Pnested}.

Since $\Acl(P,\Ks{k+1})$ is continuous in the entries of $\Ks{k+1}$, 
and since the eigenvalues of a matrix are continuous in its entries
(see, for example \cite[Theorem~5.2. on p.~89]{Serre2010}),
it follows that by a sufficiently small perturbation made to
$\Ks{k+1}$, 
along with \eqref{eq:pf_THMexists},
we still have $\lambda \notin \eigCLl{\Khs{k+1}}$.
It then follows from \eqref{eq:lambdanotinA}
and \eqref{eq:eigsubh} 
that $\lambda \notin \eigCLrh{\Khs{\star}}$.

Thus we have been able to show that there exists a
$(\Khs{k},\Khs{\star}) \in \sK{\text{RHS}}$ such that 
$\lambda \notin \eigCLrh{\Khs{\star}}$,
which contradicts our assumption. 
\end{IEEEproof}

Now we are ready to prove our main inductive step: that given a
certain number of controller indices which are allowed to be dynamic,
and the associated set of fixed modes, allowing one additional index
to become dynamic does not change the fixed modes.
\begin{thm} \label{thm:FMsksub}
\begin{onehalfspace}
Given an FDLTI
plant $P$, a sparsity pattern $\S$, 
an admissible set of dynamic elements
at step $k$ denoted by $\sIk{k} \subset \Adm{\S}$, an index pair
$(i, j) \in \Adm{\S}\setminus\sIk{k}$ 
that is further allowed to be dynamic at step $k+1$, 
and the resulting 
$\sIk{k+1} = \sIk{k} \cup (i, j)$, we have:
\end{onehalfspace}
\vskip -5pt
\begin{equation*} \label{eq:FMsksub}
\FM{P}{\S}{\Tsk{k}} ~=~ \FM{P}{\S}{\Tsk{k+1}}.
\end{equation*}
\end{thm}
\vskip 5pt
\begin{IEEEproof}
Beginning with the quantity on the right-hand side, we get:
\begin{equation*}
\begin{aligned}
\FM{P}{\S}{\Tsk{k+1}} %
& \overset{\text{Lem.} \ref{lem:indCLeq} }{=} \bigcap\limits_{\Ks{k}\in\S
\cap\Tsk{k}} \FM{\Pk{k}}{\S}{\Tspk{1}_{i, j}}
\\
&\overset{\text{Thm.} \ref{thm:FMs1eq} }{=} 
\bigcap\limits_{\Ks{k}\in\S
\cap\Tsk{k}}  \FM{\Pk{k}}{\S}{\Ts} \\
&\hskip 5pt=~ \FM{P}{\S}{\Tsk{k}}, %
\end{aligned}
\end{equation*}
where the final %
equality follows since clearly
\mbox{$(\S\cap\Tsk{k})+(\S\cap\Ts)=\S\cap\Tsk{k}$,}
and this completes the proof.
\end{IEEEproof}

We can now state and easily prove our main result.  The following
shows that for any FDLTI plant $P$, and any sparsity pattern $\S$, the set of fixed
modes with respect to static and dynamic controllers are the same.
\begin{thm} \label{thm:statdynFM}
Given plant $P$, and sparsity constraint $\S$:
\begin{equation} \label{eq:thmstatdynFM}
\FM{P}{\S}{\Ts} = \FM{P}{\S}{\Td}.
\end{equation}
\end{thm}
\begin{IEEEproof}
This follows by induction
 from Theorem~\ref{thm:FMsksub}.
\end{IEEEproof}

\subsection{Stabilization}
\label{sec:mainSuff}
The results from the previous subsection tell us that having all of
the fixed modes in the LHP is necessary for stabilizability with
respect to FDLTI controllers with the given structure. 
We now address the sufficiency of the condition.
With a constructive proof, we will show that we can stabilize 
a plant $P$ with arbitrary information structure $\S$, as long as it
has no unstable fixed modes.
We will achieve this by showing that we can always find a controller
which will reduce the number of unstable modes, while leaving all of
the fixed modes in the LHP, which can then be applied as many times as
required.

We will first state the following lemma from \cite{wang_1973},
which gives some properties regarding continuity and topology 
of non-fixed modes with respect to static controllers.
It tells us that we can keep the modes within a given distance of the
original ones by closing a small enough matrix $D$ around the plant,
and that an arbitrarily small $D$ can move all of the non-fixed modes.
\begin{lem}
\label{lem:Kexists}
For any plant $P$, and any information structure~$\S$,
partition the open-loop eigenvalues of $P$ as in \eqref{eq:eigPpart},
then we have:
\begin{enumerate}
\item For all $\epsilon > 0$, there exist $\gamma > 0 $
such that for all $D \in \S \cap \Ts$ with 
$\norm{D}_\infty < \gamma$,  there are exactly 
$\mu(\lambda,P)$ eigenvalues of $\Acl(P,D)$ in 
$B(\lambda,\epsilon)$, 
for all $\lambda \in \Lambt(P)$. 
\label{lem:samenumeig}
\item For all $\gamma > 0$, there exist
 $D \in \Ts \cap \S$ with $\norm{D}_\infty < \gamma$, such that
 $\lambda \notin \eig{\Acl(P,D)}$, 
 for all $\lambda \in \Lambt(P)$.
 \label{lem:eigchange}
\end{enumerate}
\end{lem}

\begin{IEEEproof}
See Lemma 4 in \cite{wang_1973}. 
The proof was developed for strictly proper plants with diagonal information structure,
however, it does not use any property specific to only
block-diagonal information structure and thus could be replaced
by any arbitrary information structure. 
To generalize it for the proper plants, a similar change of variable technique 
as in proof of Lemma~\ref{lem:DkSc} can be used, 
which would add an invertability constraint that almost always holds. 
\end{IEEEproof}

\begin{rem} \label{rem:randomD}
It follows from the proof that 
the set of $D$ which violate  
part \ref{lem:eigchange} of Lemma~\ref{lem:Kexists},
forms a subset with zero Lebesgue measure,
and thus a random $D \in \S$ that is sufficiently small
satisfies all of the conditions of Lemma \ref{lem:Kexists}.
Precisely, the space of static controllers that does not move the non-fixed modes 
is constructed by a finite union of hyper-surfaces in ${(\Ts \cap \S) \subset \mathbb{R}^{n_u\times n_y}}$.
Thus, a $D$ that satisfies all of the conditions of 
Lemma~\ref{lem:Kexists},
can be found with probability one by randomly choosing 
the direction of $D \in \Ts \cap \S$,
 and then scaling
it appropriately such that $\norm{D}_\infty < \gamma$.
\end{rem}

\begin{figure} [htbp]
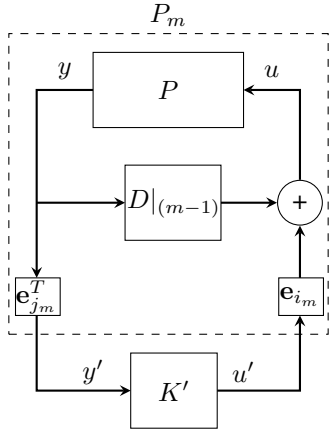
 
\begin{center}
\PDKprime
\end{center}
\caption{$P_m$ is the SISO map from $u^\prime$ to $y^\prime$, and
  $\mK$ is the map from $y$ to $u$, giving the total control for the
  original plant.} 
\label{fig:PDKp}
\end{figure}

We now establish the following theorem,
which shows how a given non-fixed mode can be extracted as a
controllable and observable mode of a specific SISO system, 
as illustrated in Figure~\ref{fig:PDKp}.
\begin{thm}\label{thm:lamijstab}
For any plant $P$ with 
$\lvert \Lambp(P) \rvert \geq 1$, 
and all fixed modes in the LHP (i.e., $\FM{P}{\S}{\Ts} \subset \LHP$),
there exists a $\Dk \in \Ts \cap \S$,
and an integer $m \in \{1,\cdots,a\}$,
such that
if we define the following SISO system:
\begin{equation}\label{eq:thmPm}
\begin{aligned}
P_m &= \e{j_m}^T \LFT{P}{\Dk} \e{i_m} =
\left[ \begin{array}{c | c}
A_m & B_m \\ \hline
C_m & D_m
\end{array} \right] \\
& \triangleq \left[ \begin{array}{c | c}
\Ap + \Bp M\Dk\Cp & \Bp M\e{i_m} \\ \hline
\e{j_m}^T N\Cp & \e{j_m}^T\Dp M \e{i_m}
\end{array} \right],
\end{aligned}
\end{equation}
the following then hold:
\begin{enumerate}
\item \label{thmen:forsome} 
 There exists $\alpha \in \Lambp(P)$,
such that $\alpha$ 
is a controllable and observable mode of $P_m$;
\item \label{thmen:unstabeigcount} 
The total number of unstable modes of $P_m$ is no greater than
that of $P$, i.e.,
$\nu(P_m) \leq \nu(P)$.
\end{enumerate}

\end{thm}

\begin{IEEEproof}
The outline of the proof is as follows.
We will first find a~$D \in \S \cap \Ts$ that when closed around~$P$,
will move all of its non-fixed modes,
and will identify the index $m \in \{1, \cdots, a\}$
for which $\Di{m}$ is the first in the sequence to alter all of them.
This means that only changing the 
$(i_m,j_m)^{\mathrm{th}}$ element of the static controller %
will change unstable mode(s) of the closed-loop,
and thus those modes must be in the controllable and observable 
modes of the SISO plant from $u_{i_m}$ to $y_{j_m}$.

Proof of part \ref{thmen:forsome}: 
Since~$\Lambp(P)\subseteq\Lambt(P)$, Lemma~\ref{lem:Kexists} 
guarantees that we can take the static gain~$D\in\S\cap\Ts$ 
such that when closed around~$P$,
would move all of its unstable non-fixed modes.
It also asserts that by choosing this~$D$ small enough,
the closed loop~$\Acl(P, D)$ would have no more unstable modes than~$P$ itself.

 Construct a sequence of matrices $\Di{m} \in \Ts \cap \S$ as in
 \eqref{eq:DmDef}, so that $\Di{a} = D$ and $\Di{0} = 0$, thus:
\begin{equation*}
\begin{aligned}
\forall \:\: \alpha \in \Lambp(P):&
\qquad \alpha \notin \eig{\Acl(P,\Di{a})},  \\
\forall \:\: \alpha \in \Lambp(P):&
\qquad \alpha \in \eig{\Acl(P,\Di{0})}. %
\end{aligned}
\end{equation*}

By decreasing $m$ from $a$ to 1, 
there must exist a value of $m \in \{1,\cdots,a\}$,
such that:
\begin{subequations} \label{eq:alphaandeig}
\begin{align} 
\forall \:\: \alpha \in \Lambp(P): &
\qquad \alpha \notin \eig{\Acl(P,\Di{m})},
\label{eq:alphaeig} \\
\exists \:\: \alpha \in \Lambp(P): &
\qquad \alpha \in \eig{\Acl(P,\Di{m-1})}; \label{eq:alphanoteig} 
\end{align}
\end{subequations}
that is, $m$ is the first index for which all of the unstable
non-fixed modes have been moved.
If we then set $\Dk = \Di{m-1} $ and use 
the definitions from \eqref{eq:thmPm}, 
as illustrated in Figure \ref{fig:PDKp},
similar to \eqref{eq:pf_LEM_moveD} we have:
\begin{equation} \label{eq:CLeigEQ}
\begin{aligned}
\Acl(P, \Di{m})~=~& \Acl(P, \Di{m-1} + \e{i_m}D_{i_m, j_m}\e{j_m}^T)
\\~\overset{\eqref{eq:LFTprop}}{=}~& 
\Acl(\LFT{P}{\Di{m-1}}, \e{i_m}D_{i_m, j_m}\e{j_m}^T) 
\\~=~& \Acl(\e{j_m}^T\LFT{P}{\Di{m-1}}\e{i_m}, D_{i_m, j_m})
\\ ~=~& \Acl(P_m, D_{i_m, j_m}).
\end{aligned}
\end{equation}
From \eqref{eq:alphanoteig}, 
 there exists at least one 
$\alpha \in \Lambp(P)$
such that:
\begin{equation*}
\alpha ~\in~ \eig{\Acl(P,\Di{m-1})} ~=~  \eig{A_m},
\end{equation*}
but due to \eqref{eq:alphaeig},
\begin{equation*}
\alpha ~\notin~ \eig{\Acl(P,\Di{m})} ~\overset{\eqref{eq:CLeigEQ}}{=}~ \eig{\Acl(P_m,D_{i_m,j_m})}.
\end{equation*}
For all such $\alpha$ that are thus moved by only closing
 $D_{i_m,j_m}$ around the SISO system $P_m$ 
(for which the only information structure is the centralized one, 
$\Sc$), we have:
\begin{equation*}
\begin{aligned}
\exists \;\; D_{i_m,j_m}\in \mathbb{R} \:\:
\mathrm{ s.t.: } \:\: 
& \alpha \notin \eig{\Acl(P_m,D_{i_m,j_m})}
\\
\Rightarrow~ & \alpha \notin \FM{P_m}{\Sc}{\Ts}.
\end{aligned}
\end{equation*}
Finally, due to Lemma \ref{lem:FM_ScTs},
the fixed modes of any FDLTI plant with centralized 
information structure are equal to its 
unobservable or uncontrollable modes, we must have that those 
$\alpha$ are controllable and observable modes of $P_m$. 

Proof of part \ref{thmen:unstabeigcount}:
since~$A_m = \Acl(P, \Di{m-1})$, 
we need to show that  
this~$\Di{m-1}$ satisfies  Lemma~\ref{lem:Kexists}.\ref{lem:samenumeig}
when we take the~$\epsilon$-balls in Lemma~\ref{lem:Kexists}
small enough such that they do not intersect with~$\RHP$.
However this is the case since 
the given~$D$ in part~\ref{thmen:forsome} of the proof satisfies 
Lemma~\ref{lem:Kexists}, 
and~$\Di{m}$ that are constructed from this~$D$,
satisfy~$\norm{\Di{m}}_\infty \leq \norm{D}_\infty \leq \gamma$ 
for any~$m\in\{0, 1,\cdots, a\}$
based on the definition.
\end{IEEEproof}

In the following proposition, %
we will use observer-based pole placement for a centralized information structure
to show how one can stabilize unstable, non-fixed
modes of~$P_m$ in~\eqref{eq:thmPm}.
We will add one further design constraint that the unstable modes of
the controller would be different than that of $P_m$, 
and will show that this constraint is always achievable 
by a small perturbation of the gains.
This ensures that an induction-based argument can be used later on.
This constraint is not mentioned in \cite{wang_1973},
and it is unclear that without such a constraint 
how one can guarantee that a rigorous induction could follow, 
even for a diagonal information structure.

\begin{prop} \label{plm:stabij}
All of the controllable and observable unstable modes of 
the plant $P_m$ can be stabilized by an observer-based controller
$\pK$ such that:
 \begin{equation} \label{eq:nullpkintersect}
\Eigp{\pK} \cap  \Eigp{\LFT{P_m}{\pK}} ~=~ \varnothing .
 \end{equation}
\end{prop}

\begin{IEEEproof}
Our proof is in a constructive manner, 
we will first find a $\pK$ to only stabilize the controllable and 
observable modes of $P_m$ 
without considering~\eqref{eq:nullpkintersect}.
We will then show that~\eqref{eq:nullpkintersect} is not satisfied 
only on a set with zero measure, 
and thus almost any small perturbation in the specific elements of~$\pK$ 
will satisfy~\eqref{eq:nullpkintersect}.

First find a similarity transformation $T$ that will put $P_m$ 
in its Kalman canonical form, therefore we would have:
\begin{equation} \label{eq:Kalman_dc2}
\begin{aligned}
\begin{bmatrix}
T & 0 \\
0 & I
\end{bmatrix}
\left[
\begin{array}{l | r}
A_m & B_m \\ \hline
C_m & D_m
\end{array}
\right]
\begin{bmatrix}
T^{-1} & 0 \\
0 & I
\end{bmatrix}
= \\
\left[
\begin{array}{l c c c | r}
\At{11}^m & 0 & \At{13}^m & 0 & \Bt{1}^m \\
\At{21}^m & \At{22}^m & \At{23}^m & \At{24}^m & \Bt{2}^m \\
0 & 0 & \At{33}^m & 0 & 0 \\
0 & 0 & \At{43}^m & \At{44}^m & 0 \\ \hline
\Ct{1}^m & 0 & \Ct{2}^m & 0 & D_m
\end{array}
\right],
\end{aligned}
\end{equation}
where as before, all the~$\tilde{(\cdot)}$ parameters depend on 
the transformation matrix $T$ and the state-space representation of~$P_m$. 
We want to stabilize all the unstable modes in $\At{11}$.
Since based on definition 
$(\At{11},\Bt{1})$ is a controllable pair and 
$(\At{11},\Ct{1})$ is an observable pair,
there exists a state feedback gain $F$ and an observer gain $L$,
such that eigenvalues of $\At{11} -	\Bt{1}F$ and 
$\At{11} -	L\Ct{1}$ can be arbitrary assigned, 
and hence can be stabilized. 
We will now show that the following controller 
will stabilize all the unstable modes of~$\At{11}$.
Take the controller as:
\begin{equation*}\label{eq:pKassign}
\pK = \left[\begin{array}{c | c}
\pA & \pB \\ \hline
\pC & 0
\end{array}
\right] = \left[
\begin{array}{c | c}
\At{11} - \Bt{1}F - L\Ct{1} + L D_m F & L \\ \hline
-F & 0
\end{array}
\right]
\begin{array}{c c c}
\end{array};
\end{equation*}
apply~$T$ from~\eqref{eq:Kalman_dc2} on~$P_m$
and close~$\pK$ around it, 
then the closed-loop~$\Acl(P_m,\pK)$ would be:
\begin{equation*} \label{eq:AclPMpK}
\begin{array}{l}
\left(
\begin{array}{l c c c  c}
\At{11} & 0 & \At{13} & 0 & -\Bt{1}F \\
\At{21} & \At{22} & \At{23} & \At{24} & -\Bt{2}F \\
0 & 0 & \At{33} & 0 & 0 \\
0 & 0 & \At{43} & \At{44} & 0 \\ 
L\Ct{1} & 0 & L\Ct{2} & 0 &  \At{11} - \Bt{1}F - L\Ct{1}
\end{array}
\right)
\end{array}
\end{equation*}
Apply another similarity transformation $T_1$, 
which keeps the first four rows the same and subtract the first row from the fifth,
then we have:
\begin{equation*} \label{eq:eigMAclPMpK}
\begin{aligned}
&\eig{\Acl(P_m,\pK)} = \eig{T_1\Acl(P_m,\pK)T_1^{-1}} = \text{eig} \\ 
&\left(
\begin{array}{l c : c c  c}
\At{11}-\Bt{1}F  & 0 & \At{13} & 0 & -\Bt{1}F \\
\At{21}-\Bt{2}F & \At{22} & \At{23} & \At{24} & -\Bt{2}F\\
\hdashline
0 & 0 & \At{33} & 0 & 0 \\
0 & 0 & \At{43} & \At{44} & 0 \\ 
0 & 0 & L\Ct{2}-\At{13} & 0 &  \At{11} - L\Ct{1}
\end{array}
\right)
\end{aligned}
\end{equation*}
Thus the eigenvalue of the closed loop would be 
\begin{equation*}
\begin{array}{l}
\eig{\Acl(P_m,\pK)} = \\ \eig{\At{11} - \Bt{1}F} \cup \:
\eig{\At{11} - L\Ct{1}} \cup \: 
\left( \bigcup\limits_{i=2}^4 \eig{\At{ii}} \right)
\end{array}
\end{equation*}
Therefore for all observer-based controllers
that naturally satisfy 
$\eig{\At{11} - \Bt{1}F} \in \LHP$ 
and $\eig{\At{11} - L\Ct{1}} \in \LHP$;
unstable modes of $\LFT{P_m}{\pK}$ would be 
independent of $F$ and~$L$, i.e.: %
\begin{equation} \label{eq:unstabPmpK}
\Eigp{(\LFT{P_m}{\pK}} ~=~ \bigcup\limits_{i=2}^4
\Eigp{\At{ii}},
\end{equation}
and all unstable modes in $\At{11}$ can be stabilized by 
appropriate choice of matrices $F$ and $L$.

We will now show that~\eqref{eq:nullpkintersect} is not met on a set with zero measure
in the ambient space of~$L$.
Replacing~\eqref{eq:unstabPmpK} in~\eqref{eq:nullpkintersect} yield that
constraint~\eqref{eq:nullpkintersect} is met if and only if:
\begin{equation} \label{eq:notmet}
\Eigp{\pK} \: \: \bigcap \: \:  \left( \bigcup\limits_{i=2}^4
\Eigp{\At{ii}} \right) ~=~ \varnothing,
\end{equation}
and if not, we enforce~\eqref{eq:nullpkintersect} 
by appropriately perturbing the~$L$ matrix. 
Construct the perturbed controller~$\hat{\pK}$ 
by replacing~$L$ in~$\pK$ with~$\hat{L} = L + L_\epsilon$, i.e.:
\begin{equation*} \label{eq:pKhatDefn}
\hat{\pK} \triangleq \left[
\begin{array}{c | c}
\hat{\pA} & \hat{L} \\ \hline
-F & 0
\end{array} \right],
\end{equation*}
with~$\hat{\pA} \triangleq \At{11} - \Bt{1}F - \hat{L}\Ct{1} + \hat{L}D_mF$.
We want to show that~$\hat{\pK}$ satisfies~\eqref{eq:notmet}
for almost any~$L_\epsilon$.
To see this, first define~$\Pp$ as:
\begin{equation*} \label{eq:PpDefn}
\Pp \triangleq \left[
\renewcommand{\arraystretch}{1.4}
\begin{array}{c | c}
\pA & I \\ \hline
- \Ct{1} + D_mF & 0
\end{array}
\right].
\end{equation*}
It is also straightforward to verify that~$\Acl(\Pp, L_\epsilon) = \hat{\pA}$. 
We want to apply Remark~\ref{rem:almostanyDk} on~$\Pp$ to show that 
almost any perturbation~$L_\epsilon$ moves all the unstable open-loop modes of~$\Pp$ 
(which is equivalent to the unstable modes of~$\pK$ as~$\eig{\Pp} = \eig{\pK}$).
This would be achieved by showing 
that non of the unstable modes of~$\Pp$ would be a fixed one, 
precisely:
\begin{equation*}
\begin{array}{l}
\Acl(\Pp, -L)~=~ \At{11}-\Bt{1}F 
\\ \Rightarrow  \FM{\Pp}{\Sc}{\Ts} \subseteq \eig{\At{11} - \Bt{1}F} \subset \LHP,
\end{array}
\end{equation*}
as~$F$ is chosen to stabilize~$\At{11}$.
Moreover, given that~$\eig{\At{11} - L\Ct{1}}\subset \LHP$,
if we chose~$L_\epsilon$ sufficiently small, 
then due to a continuity argument
we have~$\eig{\At{11} - \hat{L}\Ct{1}} \subset \LHP$.
Thus any sufficiently small perturbation~$L_\epsilon$ 
will make~$\hat{\pK}$ satisfy~\eqref{eq:nullpkintersect} 
while still keeping~$\At{11}  - \hat{L}\Ct{1}$ stable.
\end{IEEEproof}

We will encapsulate the desired properties of the intermediate controller at each step
that partially stablizes the plant in the following corollary, 
which combines Theorem~\ref{thm:lamijstab} and Proposition~\ref{plm:stabij}.
\begin{cor} \label{cor:dynij}
For every plant $P$ that satisfies the assumptions of 
Theorem \ref{thm:lamijstab}, 
there exists an~$m\in\{1,\cdots,a\}$
and a controller~$\mK  \in \S\cap\Tsp_{i_m,j_m}$ such that:
\begin{align}
&\nu(\LFT{P}{\mK}) ~\leq~ \nu(P) \: -1, \label{eq:countunstabeigs} \\
&\Eigp{\mK} \cap  \Lambp(\LFT{P}{\mK})  ~=~ \varnothing. \label{eq:nullCLKm}
\end{align}
\end{cor}
\begin{IEEEproof}
Use Theorem \ref{thm:lamijstab} 
to find $\Dk$ and $m$, use %
Proposition \ref{plm:stabij} to
find $\pK$, and construct the MIMO controller 
$\mK \triangleq \Di{m-1} + \e{i_m}\pK\e{j_m}^T$. 
As illustrated in Figure \ref{fig:PDKp},
this $\mK$ has the following
state-space representation:
\begin{equation} \label{eq:ssKm}
\mK = 
\left[
\begin{array}{c | c}
\mA & \mB \\ \hline
\mC & \mD
\end{array}
\right]
=
\left[
\begin{array}{c | c}
\pA & \pB \e{j_m}^T \\ \hline
\e{i_m} \pC & \Dk
\end{array}
\right],
\end{equation}
and clearly satisfies:
\begin{equation} \label{eq:eqAclPmPmK}
\Acl(P_m,\pK) = \Acl(P,\mK).
\end{equation}
Due to Theorem \ref{thm:lamijstab} and Proposition 
\ref{plm:stabij}, $\pK$ will stabilize at least
one unstable mode of $P$, hence we have 
$\nu(\LFT{P_m}{\pK}) \leq \nu(P)  -1$,
and thus~\eqref{eq:countunstabeigs}
would be an immediate result of this property of  $\pK$  
combined with \eqref{eq:eqAclPmPmK}.
Finally, \eqref{eq:nullCLKm} follows from~\eqref{eq:nullpkintersect}
as~$\mA = \pA$ and~$\Lambp(\LFT{P_m}{\pK}) = \Lambp(\LFT{P}{\mK})$, 
due to~\eqref{eq:ssKm} and~\eqref{eq:eqAclPmPmK}.
\end{IEEEproof}

We use induction to prove that if
all the fixed modes of $P$ are in LHP,
then we can stabilize $P$ by dynamic controller. 
We will first define the following interconnection 
that will be useful in the induction. 
Let $\Gs{0} \triangleq P$ and at each step $k$,
denote the transfer function from $u$ to $y$,
as illustrated in Figure \ref{fig:Gnested}, by $\Gs{k+1}$, i.e., 
$\Gs{k+1} = \LFT{\Gs{k}}{\Kms{k}}$. 
Let $(\Ags{k},\Bgs{k},\Cgs{k},\Dgs{k})$ be a state-space representation 
for $\Gs{k}$, 
also denote the total number of unstable modes of $\Gs{k}$ by 
$\nuk{k}$: %
\begin{equation*} \label{eq:nukdef}
\nuk{k} \triangleq \sum\limits_{\alpha \in \Lambp(\Gs{k})}  
\mu(\alpha,\Gs{k}).
\end{equation*}

\begin{figure} [htbp] 
\begin{center}
\Gnested
\end{center}
\caption{Plant $\Gs{k+1} \triangleq \LFT{\Gs{k}}{\Kms{k}}$.} 
\label{fig:Gnested}
\end{figure}

The induction will be in such a way that in each step $k$, 
we will find an integer $\matk \in \{1,\cdots,a \}$, and a
$\Kms{k} \in \Tsp_{i_{\matk},j_{\matk}} \cap \S$
that when closed around $\Gs{k}$,
will stabilize at least one unstable mode of $\Gs{k}$, 
thus $\nuk{k+1} \leq \nuk{k} - 1$. 
Then we will treat the corresponding $\Gs{k+1}$ as 
the new plant for which we want to stabilize 
the rest of remaining $\nuk{k+1}$ unstable eigenvalues, thus in at most
$\nuk{0}$
steps, $P$ will be stabilized. 
A crucial part of induction is that $\Gs{k+1}$
must have no fixed mode in closed RHP, 
this is not addressed in \cite{wang_1973} and at this point 
it is directly claimed that Theorem \ref{thm:FMstab} holds true.
We will formalize this fact with the help of following lemma.
It is enough to show that closing $\mK$ around $P$ does not add
any unstable fixed modes to $\LFT{P}{\mK}$. 
\begin{lem} \label{lem:GsFM}
Assume that all the fixed modes of $P$ are
in LHP, i.e.:
\begin{equation} \label{eq:GkLHP}
\FM{P}{\S}{\Ts} \subset \LHP,
\end{equation}
also, assume that a controller $\mK$ 
is such that it satisfies \eqref{eq:nullCLKm},
then we have:
\begin{equation*} \label{eq:GkpLHP}
\FM{\LFT{P}{\mK}}{\S}{\Ts} \subset \LHP.
\end{equation*}
\end{lem}
\begin{IEEEproof}
Proof is done by contradiction, we will first 
create the following set-up to state the idea.
Let $(\Ak,\Bk,\Ck,\Dk)$
be a state-space representation for $\mK$. 
We have:
\begin{equation*} \label{eq:FMPinFMPKm}
\FM{P}{\S}{\Ts} \subseteq \FM{\LFT{P}{\mK}}{\S}{\Ts},
\end{equation*}
since the RHS is the set of fixed modes with respect to controllers in the form~$\mK + \S\cap\Ts$, 
whereas the LHS equals~$\FM{P}{\S}{\Td}$ (by Theorem~\ref{thm:statdynFM}), 
that is the set of fixed modes with respect to controllers in~$\S\cap\Td$, 
which is a bigger set than $\mK\in \S\cap\Td$.
Next, it is trivial to check that if we close $-\mK$ around 
$\LFT{P}{\mK}$, then by applying a similarity transformation 
$T_2$, 
a state-space realization that does not omit non-minimal modes
of~$\LFT{\LFT{P}{\mK}}{-\mK}$ can be written as:
\begin{equation} \label{eq:LFTGkpKm}
\begin{aligned}
&\begin{bmatrix}
T_2 & 0 \\
0 & I
\end{bmatrix}
\LFT{\LFT{P}{\mK}}{-\mK}
\begin{bmatrix}
T_2^{-1} & 0 \\
0 & I
\end{bmatrix}\\
&\hskip 30pt = 
\left[
\begin{array}{c c : c | c}
\Ap & \Bp\Ck & 0 & \Bp \\
0 & \Ak & 0 & 0 \\ \hdashline 
\Bk\Cp & \Bk\Dp\Ck & \Ak & \Bk\Dp \\ \hline
\Cp & \Dp\Ck & 0 & \Dp
\end{array}
\right],
\end{aligned}
\end{equation}
thus we have 
\begin{equation*}
\eig{\LFT{\LFT{P}{\mK}}{-\mK}} = \eig{\Ap} \cup 
\eig{\Ak}.
\end{equation*}
Furthermore, due to \eqref{eq:GkLHP}, there exist a 
$D\in\Ts \cap \S$ that will move all the unstable modes of~$\Ap$.
If we apply the same~$D$ on \eqref{eq:LFTGkpKm}, 
due to the block-diagonal structure we have
$\eig{\Acl(\LFT{\LFT{P}{\mK}}{-\mK}, D)} = \eig{\Acl(P, D)} \cup \eig{\Ak}$, 
which yields: %
\begin{equation}\label{eq:FMrelations}
\begin{aligned}
\FM{\LFT{P}{\mK}}{\S}{\Ts}  %
\overset{}{\subseteq}
\FM{P}{\S}{\Ts} \cup \eig{\Ak}.
\end{aligned}
\end{equation}

Now we are ready to do the main contradiction part, assume that 
there exist an $\alpha \in \FM{\LFT{P}{\mK}}{\S}{\Ts}$, 
with $\Re(\alpha) \geq 0$, then
\begin{equation*}
\begin{aligned}
\alpha& \in
\FM{\LFT{P}{\mK}}{\S}{\Ts}, \qquad \Re(\alpha) \geq 0 \\
\alpha &
 \overset{\eqref{eq:FMrelations}}{\in} 
\FM{P}{\S}{\Ts} \cup \eig{\Ak} \\
\overset{\eqref{eq:GkLHP}}{\Rightarrow}
\alpha& \in  \eig{\Ak} \\
\overset{ \eqref{eq:nullCLKm} }{\Rightarrow}
\alpha & \notin  \eig{\LFT{P}{\mK}} \\ %
\Rightarrow 
\alpha & \notin \FM{\LFT{P}{\mK}}{\S}{\Ts} 
\end{aligned}
\end{equation*}
thus we have achieved the desired contradiction.
\end{IEEEproof}

Constraint~\eqref{eq:nullCLKm} in Corollary~\ref{cor:dynij} 
ensures that the unstable modes be non-overlapping, 
and is one sufficient condition to prove Lemma~\ref{lem:GsFM}.
When this condition is not met for an arbitrary choice of the feedback/observer gain, 
one way to always make it feasible is
by adding the perturbation~$L_\epsilon$ to the observer gain, 
which in turn might prevent exact pole placement.
However, one can place the poles arbitrarily close to the desired locations
by choosing~$L_\epsilon$ sufficiently small.

Now we are ready to claim that if all the fixed modes 
of $P$ are in the LHP, then we can stabilize $P$ by a dynamic controller.
This stabilizing controller would be a summation of individual controllers $\Kms{k}$, 
each obtained in one step of the induction,
where in each step $k$, 
$\Kms{k}$ would only have one dynamic element 
(i.e., $ \Kms{k} \in \Tsp_{i_{\matk},j_{\matk}} \cap \S$, for some
 $\matk \in \{1,\cdots,a\}$).
 \vspace{2 mm}

\begin{thm} \label{thm:suff}
For any FDLTI plant~$P$, and any information structure~$\S$, 
if $\FM{P}{\S}{\Ts} \subset \LHP$,
then there exist a controller 
$K \in \S \cap \Td$ that will stabilize $P$.
\end{thm}
\begin{IEEEproof}
Proof is done by induction.
Take~$k\leftarrow 0$ and let~$\Gs{0} \triangleq P$.
As per assumption of this theorem,
$\FM{\Gs{0}}{\S}{\Ts} = \FM{P}{\S}{\Ts} \subset \LHP$.
At each induction step $k$, 
we would stabilize at least one of the unstable modes of~$\Gs{k}$
by Corollary~\ref{cor:dynij}.
Specifically, with~$P$ replaced by $\Gs{k}$ in Corollary \ref{cor:dynij}, 
we can find a $\matk \in \{1,\cdots,a\}$, and a controller 
$\Kms{k}\in \S \cap \Tsp_{i_{m^{(0)}},j_{m^{(0)}}}$, that will 
stabilize at least one of unstable modes of $\Gs{k}$. %
This~$\Kms{k}$ satisfies~\eqref{eq:nullCLKm} 
(with $P$ replaced by $\Gs{k}$),
and thus by Lemma \ref{lem:GsFM},
$\Gs{k+1} = \LFT{\Gs{k}}{\Kms{k}}$, would have all of its
fixed modes in LHP, i.e., $\FM{\Gs{k+1}}{\S}{\Ts} \in \LHP$. 
This guarantees that we can proceed with the induction by taking $k\leftarrow k +1$, 
as long as~$\Gs{k}$ has any remaining unstable mode. 
Since at each step at least one unstable mode is stabilized,
$P$ would be stabilized in at most $\nu(P)$ steps.
The final~$K \in \S \cap \Td$ that will stabilize~$P$, 
is equal to the summation of controllers at each step, i.e.:
\begin{equation*}
K(s) \overset{\eqref{eq:LFTprop}}{=} \sum\nolimits_k \Kms{k}(s).
\end{equation*}
\end{IEEEproof} 
We can easily show that stability of all the fixed modes of~$P$,
$\FM{P}{\S}{\Ts}\subset \LHP $, 
is also a necessary condition for the existence of stabilizing controller:

\begin{thm} \label{thm:FMstab}
A plant $P$ is stabilizable by a controller $K \in  \S \cap \Td$, 
if and only if $\FM{P}{\S}{\Ts}\subset \LHP$.
\end{thm}
\begin{IEEEproof}
The sufficiency part is done in Theorem \ref{thm:suff}. 
For the necessity part note that static fixed modes can not be moved by the dynamic controller either
(Theorem~\ref{thm:statdynFM}), i.e.:
\begin{equation*}
\begin{array}{l l}
& \FM{P}{\S}{\Ts} \not\subset  \LHP \\
\overset{\mathrm{Thm.} \ref{thm:statdynFM}}{\Rightarrow} &
\FM{P}{\S}{\Td} \not\subset \LHP  \\
\overset{\mathrm{by def}}{\Rightarrow} &
\nexists\:\:K \in \Td \cap \S  \quad \mathrm{ s.t. } \quad
\Acl(P,K) \subset \LHP.
\end{array}
\end{equation*}
\end{IEEEproof}

\section{Synthesis and Numerical Example}
\label{sec:numex}
In this section we provide an explicit algorithm to stabilize a plant
which has no unstable fixed modes, 
and run it on one numerical example to illustrate its 
implementation.
 Algorithm \ref{alg:FindK} is distilled from the steps 
taken in the paper to prove the main theorem, 
and thus can almost certainly be improved upon in several respects.

In Algorithm \ref{alg:FindK}, $\Dkalg$ is chosen randomly at each outer-step,
and as stated in Remark~\ref{rem:randomD},
would be a valid choice with probability one.
This $\Dkalg$ must be chosen small enough 
($\norm{\Dkalg}_\infty <\atk{\gamma}$) 
such that 
the total number of unstable modes would not increase
when each element of the sequence $\{\Dkalgm\}_{m=1}^a$ 
is closed around $\Gs{k}$.
A prior knowledge of such an upper bound on $\Dkalg$,
denoted by $\atk{\gamma}$, 
is not available and is hard to attain. 
This leads us to consider the alternative approach of
repeatedly making~$\Dkalg$ smaller in a loop until 
Theorem \ref{thm:lamijstab}.\ref{thmen:unstabeigcount} holds.
This iterative scaling repeats itself when 
\eqref{eq:nullpkintersect} is not met. In this case, 
 as proof suggests, we perturb $\atk{L}$ by $\atk{\hat{L}}$.
This perturbation must be chosen small enough that it will not
 make any modes of 
 $\atk{\At{11}} - (\atk{L} + \atk{\hat{L}}  )\atk{\Ct{1}}$ 
 unstable. The upper bound on the perturbation $\atk{\hat{L}}$ is 
 unknown, and thus, similar to the case for $\Dkalg$,
 we iterate to make it small enough to meet the constraints.

\setlength{\intextsep}{0pt} 
\begin{algorithm}[tbph]
\caption{Finding a controller $K\in \Td \cap \S$ to stabilize~$P$}
\label{alg:FindK}
\begin{algorithmic}
\REQUIRE Plant $P$, information structure $\S$
\ENSURE Controller $K\in \Td \cap \S$ that will stabilize $P$
\STATE $k \leftarrow 0$, $\Gs{0} \leftarrow P$,
$K(\sigma) \leftarrow 0$
\STATE /* Repeat the outer loop until the plant is stabilized */ 
\WHILE{$\lvert \Lambp(\Gs{k}) \rvert \geq 1$}
\STATE /* Select a static controller as in Rem.~\ref{rem:randomD} */
\STATE Choose a random $D \in \Ts \cap \S$
\WHILE {$\nu(\LFT{\Gs{k}}{D}) > \nu(\Gs{k}) $}
\STATE $D \leftarrow D/2 $
\ENDWHILE
\STATE /* Find a controllable index as in Thm.~\ref{thm:lamijstab} */
\STATE $\matk \leftarrow a$
\WHILE{ $\Lambp(\LFT{\Gs{k}}{\Di{\matk -1} })
\cap \Lambp(\Gs{k}) = \varnothing$} 
\STATE $\matk \leftarrow \matk -1$
\ENDWHILE
\STATE /* Form the SISO plant as in Fig.~\ref{fig:PDKp} */
\STATE $\Gs{k}_{\matk} \leftarrow  
 \e{j_{\matk}}^T \LFT{\Gs{k}}{\Di{\matk -1}} \e{i_{\matk}}$
\STATE /* Stabilize the SISO plant as in Prop.~\ref{plm:stabij} */
\STATE Find a Kalman similarity transformation $\atk{T}$ for 
$\Gs{k}_{\matk}$
\STATE Name all the corresponding partitions by 
$\atk{(\tilde{\cdot})} $
\STATE Find a $\atk{F}$ to stabilize 
$\atk{\At{11}} - \atk{\Bt{1}}\atk{F}$
\STATE Find a $\atk{L}$ to stabilize 
$\atk{\At{11}} - \atk{L}\atk{\Ct{1}}$
\STATE/* Ensuring that constraint~\eqref{eq:nullCLKm} holds */
\STATE $\atk{M} \leftarrow (I - \Di{\matk-1}\Dgs{k})^{-1}$
\STATE $\atk{\Dshort} \leftarrow \e{j_{\matk}}^T \Dgs{k}\atk{M} \e{i_{\matk}}$
\IF{$ \Eigp{\atk{\At{11}} - \atk{\Bt{1}}\atk{F} + 
\atk{L}(\atk{\Dshort} \atk{F} -\atk{\Ct{1}} ) } \cap 
\left(
\bigcup\limits_{i=2}^4 \Eigp{\atk{\At{ii}}}
\right)
\neq \varnothing
$}
\STATE /* Perturb the observer gain if~\eqref{eq:nullCLKm} does not hold */
\STATE Choose a random $\atk{L}_\epsilon$ 
\STATE /* Make the perturbation sufficiently small not to have any new unstable mode */
\WHILE{ $\lvert 
\Eigp{ \atk{\At{11}} - 
(\atk{L} + \atk{L}_\epsilon )\atk{\Ct{1}}} \rvert \geq 1 $ }
\STATE $\atk{L}_\epsilon \leftarrow \atk{L}_\epsilon/2$
\ENDWHILE
\STATE $\atk{L} \leftarrow \atk{L} + \atk{L}_\epsilon$
\ENDIF
\STATE /* Construct the MIMO controller as in Cor.~\ref{cor:dynij} */
\vspace{.25 em}
\STATE $\Ks{k} \leftarrow \left[\begin{array}{c | c}
\begin{array}{c}
\atk{\At{11}} - \atk{\Bt{1}}\atk{F} + \\
 \atk{L}\left( \atk{\Dshort}\atk{F} -\atk{\Ct{1}}\right)
\end{array} &
\atk{L} \e{j_{\matk}}^T \\ \hline
-\e{i_{\matk}}\atk{F} & \Di{\matk - 1}
\end{array}\right]$
\STATE $K \leftarrow K + \Ks{k}$
\STATE $\Gs{k+1} \leftarrow \LFT{\Gs{k}}{\Ks{k}}$
\STATE $k \leftarrow k+1$
\ENDWHILE
\RETURN $K$
\end{algorithmic}
\end{algorithm}
\begin{rem} \label{rem:algifthen}
The intersection in the \textbf{if-then} section in Algorithm~\ref{alg:FindK} 
would almost always result in a null set if interpreted with unlimited precision. 
However, choosing to replace the exact intersection with a proximity condition 
could possibly avoid very large feedback and observer gains.
\end{rem}

\begin{rem} \label{rem:LHP}
We can replace $\LHP$ throughout the paper with another open set of acceptable closed-loop
eigenvalues, letting its complement replace $\RHP$ as the closed set of
unacceptable closed-loop eigenvalues. 
The results of Section~\ref{sec:mainNecc} hold up to show that the
fixed modes must not be in the unacceptable region, the results of
Section~\ref{sec:mainSuff} hold up to show that if they are not, then
all of the modes can be moved to the acceptable region, and
Algorithm~\ref{alg:FindK} can be applied to find a controller which
achieves that objective. 
One can further define a smaller open set of desirable closed-loop
eigenvalues into which all of the non-fixed modes can be moved by
Algorithm~\ref{alg:FindK}, taking note of the possibility of fixed and
non-fixed modes overlapping in the acceptable-yet-undesirable region,
as mentioned in Section~\ref{sec:prelims}.

\end{rem}

The following numerical example will use Algorithm \ref{alg:FindK} 
to stabilize the plant $P$.
\begin{ex}
Consider the following plant:
\begin{equation*}
A = \mathrm{diag}(2,3,5,-1,-1)
\end{equation*}
\begin{equation*}
B = %
\begin{bmatrix}
0 & 0 & 3 & 0 & 2 \\
0 & 0 & 0 & 1 & 0 \\
0 & 0 & 2 & 0 & 5 \\
1 & 0 & 0 & 0 & 0 \\
0 & 2 & 0 & 0 & 0
\end{bmatrix}
\;\; C = %
\bmat{
4 & 0 & 8 & 0 & 0 \\
0 & 1 & 0 & 0 & 0 \\
6 & 0 & 3 & 0 & 0 \\
0 & 0 & 0 & 5 & 0 \\
0 & 0 & 0 & 0 & 6
}%
\end{equation*}
and $D=0$. Let the information constraint for the controller 
be given by the admissible-to-be-nonzero indices:
$\Adm{\S}=\{(1,1),(3,1),(4,1),(5,2),(1,3),(3,3),(4,3),(5,4),(5,5)\}$.
This plant has fixed mode $\FM{P}{\S}{\Ts} = \{-1\}$. 
If we follow Algorithm \ref{alg:FindK} to stabilize~$P$,
and choose our desired closed-loop modes of
$\LFT{P}{K}$ 
to be
$\begin{bmatrix}
-0.5 & -1 & -1 & -1.5 & -2 & -2.5 & -3 & -3.5
\end{bmatrix}^T$,
this is %
achieved by the following resulting controller:
\begin{equation*}
\begin{aligned}
\Ak =&  %
\begin{bmatrix}
14.9224 & -460.4053 & -4.6620 \\
0.3742 & -24.4429 & 0.7485 \\
22.9223 & -763.8424 & -25.4224
\end{bmatrix}%
\\
\Bk =& %
\left[\begin{array}{l c r}
0_{3\times 1} & 
\begin{array}{c}
317.1161 \\
27.4429 \\
405.6193 \\
\end{array} & 0_{3\times 3}
\end{array}
\right]%
 \\
\Ck =& %
\left[\begin{array}{c c}
0_{3\times 4} &
\begin{array}{c}
3.9020 \\ -71.6446 \\-7.4494
\end{array}
\end{array}
\right]^T %
\\
\Dk =& %
\left[\begin{array}{c c}
\begin{array}{c}
0.0854 \\
0 \\
0.4265 \\
0.0936 \\
0 \\
\end{array}
& 0_{5\times 4}
\end{array}
\right]%
.
\end{aligned}
\end{equation*}
An alternative approach is taken in \cite{Davison_90}, in which, 
at each step, 
a (possibly dynamic) stabilizing controller is applied at
the next diagonal element of the controller,
and it is shown that by adding stabilizing controllers at each step,
the set of (possibly unstable) fixed modes are reduced, 
until the last step where the remaining fixed modes 
must be necessarily stable.
Applying the method of \cite{Davison_90} on this plant
would result in a stabilizing controller of order 7,
as compared to 3 here. 
An explanation could be that in \cite{Davison_90}, 
a (possibly dynamic) stabilizing controller is applied 
at each of the elements, 
resulting in abundant of controller states,
whereas in here, only for each unstable mode,
a stabilizing controller (not necessarily of order~1) is needed.

If we look at each of the nine SISO maps from~$u_{i_m}$ to~$y_{j_m}$ in~$P$,
then the union of controllable and observable modes of all
these SISO maps are $\{2,5\}$, 
which does not contain the unstable mode 3.
This shows that 
a static gain (the $D_{m-1}$ of Figure~\ref{fig:PDKp}) might be
necessary to assign some modes in decentralized settings, 
which is counter-intuitive
compared to the centralized case where a stabilizing
observer-based controller would have zero static gain.
\end{ex}

\section{Conclusion}
\label{sec:conc}
We revisited, verified, and generalized classic work in the
stabilizability of decentralized systems.
We generalized the notion of fixed modes to arbitrary
information structure, and provided a rigorous inductive proof that
plant modes which cannot be moved by static LTI controllers with the
prescribed structure cannot be moved by dynamic LTI controllers either.
We addressed the placement of the modes which are not
fixed, and 
showed that they can be moved to within a chosen accuracy of any
desired pole locations, 
thus solidifying and generalizing
the main result of \cite{wang_1973}.
Combining these results, we have shown that having all fixed modes in the
LHP with respect to static LTI controllers of a given information structure is
necessary and sufficient for stabilizability by dynamic LTI
controllers with the same structure.
We lastly presented an explicit algorithm for finding such a stabilizing
decentralized controller.

\newpage
\bibliographystyle{IEEEtran}%

\bibliography{IEEEabrv,distributed_control}

\end{document}